\documentclass[twocolumn, times, trackchanges]{aastex63}  
\usepackage{longtable}
\usepackage{booktabs}

\newcommand{\Msun}{$M_{\odot}$}
\newcommand{\Mej}{$M_{\rm ej}$}
\newcommand{\EK}{$E_{\rm kin}$}
\newcommand{\Mni}{$M_{\rm Ni}$}

\newcommand{\kms}{km~s$^{-1}$}
\newcommand{\ergs}{erg s$^{-1}$}

\newcommand{\OI}{O~{\sc i}}

\newcommand{\CII}{C~{\sc ii}}
\newcommand{\NaI}{Na~{\sc i}}

\newcommand{\MgII}{Mg~{\sc ii}}

\newcommand{\SiII}{Si~{\sc ii}}
\newcommand{\SiIII}{Si~{\sc iii}}

\newcommand{\SII}{S~{\sc ii}}
\newcommand{\CaII}{Ca~{\sc ii}}

\newcommand{\FeII}{Fe~{\sc ii}}
\newcommand{\FeIII}{Fe~{\sc iii}}

\newcommand{\Cofs}{$^{56}$Co}
\newcommand{\Nifs}{$^{56}$Ni}

\newcommand{\sBV}{$s_{BV}$}

\newcommand{\mb}{${\Delta}m_{15}(B)$}

\newcommand{\Angst}{$\rm \AA$}
\newcommand{\RCSi}{$R$({C \sc ii}/{Si \sc ii})}

\newcommand{\dmvalue}{$1.02 \pm 0.07$}
\newcommand{\sbvalue}{$0.94 \pm 0.03$}
\newcommand{\distmd}{$34.00 \pm 0.09$}
\newcommand{\mbmag}{$-19.12 \pm 0.11$}
\newcommand{\mbobs}{$14.88 \pm 0.02$}
\newcommand{\ebvalue}{$0.06 \pm 0.06$}
\newcommand{\tbmax}{$58,066.64 \pm 0.36$}
\newcommand{\tzero}{$58,050.22 \pm 1.20$}
\newcommand{\tlcv}{$15.86 \pm 0.76$}
\newcommand{\tgama}{$28.37 \pm 3.44$}
\newcommand{\trise}{$18.26 \pm 0.97$}
\newcommand{\Lmax}{$1.25 \times 10^{43}$\,erg\,s$^{-1}$}
\newcommand{\MniValue}{$0.63 \pm 0.02\,M_{\odot}$}
\newcommand{\VSiII}{$9550 \pm 170$\,\kms}
\newcommand{\VSiIIDot}{$130 \pm 7$\,km\,s$^{-1}$\,d$^{-1}$}
\newcommand{\RSiII}{$0.18 \pm 0.03$}

\begin{document}

\shorttitle{Observations of SN 2017hpa}
\shortauthors{Zeng et al.}

\title{SN 2017hpa: A Nearby Carbon-Rich Type Ia Supernova with a Large Velocity Gradient}

\author{Xiangyun Zeng}
\affil{Xinjiang Astronomical Observatory, Chinese Academy of Sciences, Urumqi, Xinjiang 830011, People's Republic of China}
\affil{School of Astronomy and Space Science, University of Chinese Academy of Sciences, Beijing 100049, People's Republic of China}

\author{Xiaofeng Wang}
\email{wang\_xf@mail.tsinghua.edu.cn}
\affil{Physics Department and Tsinghua Center for Astrophysics (THCA), Tsinghua University, Beijing, 100084, People's Republic of China}
\affil{Beijing Planetarium, Beijing Academy of Science of Technology, Beijing 100044, People's Republic of China}

\author{Ali Esamdin}
\email{aliyi@xao.ac.cn}
\affil{Xinjiang Astronomical Observatory, Chinese Academy of Sciences, Urumqi, Xinjiang 830011, People's Republic of China}

\author{Craig Pellegrino}  
\affil{Department of Physics, University of California, Santa Barbara, CA 93106-9530, USA}
\affil{Las Cumbres Observatory, 6740 Cortona Drive Suite 102, Goleta, CA 93117-5575, USA}

\author{WeiKang Zheng} 
\affil{Department of Astronomy, University of California, Berkeley, CA 94720-3411, USA}

\author{Jujia Zhang}  
\affil{Yunnan Observatories (YNAO), Chinese Academy of Sciences, Kunming 650216, People's Republic of China}
\affil{Key Laboratory for the Structure and Evolution of Celestial Objects, Chinese Academy of Sciences, Kunming 650216, People's Republic of China}
\affil{Center for Astronomical Mega-Science, Chinese Academy of Sciences, 20A Datun Road, Chaoyang District, Beijing, 100012, People's Republic of China}

\author{Jun Mo}   
\affil{Physics Department and Tsinghua Center for Astrophysics (THCA), Tsinghua University, Beijing, 100084, People's Republic of China}

\author{Wenxiong Li} 
\affil{Physics Department and Tsinghua Center for Astrophysics (THCA), Tsinghua University, Beijing, 100084, People's Republic of China}
\affil{The School of Physics and Astronomy, Tel Aviv University, Tel Aviv 69978, Israel}

\author{D. Andrew Howell}  
\affil{Las Cumbres Observatory, 6740 Cortona Drive Suite 102, Goleta, CA 93117-5575, USA}
\affil{Department of Physics, University of California, Santa Barbara, CA 93106-9530, USA}

\author{Alexei V. Filippenko}  
\affil{Department of Astronomy, University of California, Berkeley, CA 94720-3411, USA}
\affil{Miller Senior Fellow, Miller Institute for Basic Research in Science, University of California, Berkeley, CA 94720, USA}


\author{Han Lin}    
\affil{Physics Department and Tsinghua Center for Astrophysics (THCA), Tsinghua University, Beijing, 100084, People's Republic of China}


\author{Thomas G. Brink} 
\affil{Department of Astronomy, University of California, Berkeley, CA 94720-3411, USA}

\author{Edward A. Baron}   
\affil{Homer L. Dodge Department of Physics and Astronomy, University of Oklahoma, USA}

\author{Jamison Burke}  
\affil{Las Cumbres Observatory, 6740 Cortona Drive Suite 102, Goleta, CA 93117-5575, USA}
\affil{Department of Physics, University of California, Santa Barbara, CA 93106-9530, USA}

\author{James M. DerKacy}   
\affil{Homer L. Dodge Department of Physics and Astronomy, University of Oklahoma, USA}

\author{Curtis McCully}  
\affil{Las Cumbres Observatory, 6740 Cortona Drive Suite 102, Goleta, CA 93117-5575, USA}
\affil{Department of Physics, University of California, Santa Barbara, CA 93106-9530, USA}

\author{Daichi Hiramatsu}  
\affil{Las Cumbres Observatory, 6740 Cortona Drive Suite 102, Goleta, CA 93117-5575, USA}
\affil{Department of Physics, University of California, Santa Barbara, CA 93106-9530, USA}

\author{Griffin Hosseinzadeh}  
\affil{Center for Astrophysics \textbar{} Harvard \& Smithsonian, 60 Garden Street, Cambridge, MA 02138-1516, USA}
\affil{Las Cumbres Observatory, 6740 Cortona Drive Suite 102, Goleta, CA 9311-5575, USA}
\affil{Department of Physics, University of California, Santa Barbara, CA 93106-9530, USA}

\author{Benjamin T. Jeffers}  
\affil{Department of Astronomy, University of California, Berkeley, CA 94720-3411, USA}

\author{Timothy W. Ross} 
\affil{Department of Astronomy, University of California, Berkeley, CA 94720-3411, USA}

\author{Benjamin E. Stahl} 
\affil{Department of Astronomy, University of California, Berkeley, CA 94720-3411, USA}
\affil{Department of Physics, University of California, Berkeley, CA 94720-7300, USA}
\affil{Marc J. Staley Graduate Fellow}

\author{Samantha Stegman}  
\affil{Department of Astronomy, University of California, Berkeley, CA 94720-3411, USA}
\affil{Department of Chemistry, University of Wisconsin, Madison, WI 53706, USA}

\author{Stefano Valenti}  
\affil{Department of Physics, University of California, Davis, CA 95616, USA}

\author{Lifan Wang}  
\affil{George P. and Cynthia Woods Mitchell Institute for Fundamental Physics \& Astronomy, Texas A\&M University, Department of Physics and Astronomy, 4242 TAMU, College Station, TX 77843, USA}

\author{Danfeng Xiang}   
\affil{Physics Department and Tsinghua Center for Astrophysics (THCA), Tsinghua University, Beijing, 100084, People's Republic of China}

\author{Jicheng Zhang} 
\affil{Physics Department and Tsinghua Center for Astrophysics (THCA), Tsinghua University, Beijing, 100084, People's Republic of China}

\author{Tianmeng Zhang} 
\affil{Key Laboratory of Optical Astronomy, National Astronomical Observatories, Chinese Academy of Sciences, Beijing 100012, People's Republic of China}


\begin{abstract}
We present extensive, well-sampled optical and ultraviolet photometry and optical spectra of the Type Ia supernova (SN~Ia) 2017hpa. The light curves indicate that SN 2017hpa is a normal SN~Ia with an absolute peak magnitude of $M_{\rm max}^{B} \approx$ \mbmag\ mag and a post-peak decline rate \mb\ = \dmvalue\ mag. According to the quasibolometric light curve, we derive a peak luminosity of \Lmax and a \Nifs\ mass of \MniValue. The spectral evolution of SN 2017hpa is similar to that of normal SNe~Ia, while it exhibits unusually rapid velocity evolution resembling that of SN 1991bg-like SNe~Ia or the high-velocity subclass of SNe~Ia, with a post-peak velocity gradient of $\sim$ \VSiIIDot. Moreover, its early spectra ($t < -7.9$\,d) show prominent \CII~$\lambda$6580 absorption feature, which disappeared in near-maximum-light spectra but reemerged at phases from $t\,\sim\,+8.7$ d to $t\,\sim\,+11.7$ d after maximum light. This implies that some unburned carbon may mix deep into the inner layer, and is supported by the low \CII~$\lambda$6580 to \SiII~$\lambda$6355 velocity ratio ($\sim 0.81$) observed in SN~2017hpa. The \OI~$\lambda$7774 line shows a velocity distribution like that of carbon. The prominent carbon feature, low velocity seen in carbon and oxygen, and large velocity gradient make SN 2017hpa stand out from other normal SNe~Ia, and are more consistent with predictions from a violent merger of two white dwarfs. Detailed modelling is still needed to reveal the nature of SN 2017hpa.
\end{abstract}


\section{Introduction}

Type Ia supernovae (SNe~Ia) are widely believed to arise from explosions of carbon-oxygen (CO) white dwarfs (WDs) in a binary system, which have a typical absolute $V$-band peak magnitude of $\sim -19$\,mag \citep{phi93,per99,wxf06}. The relatively uniform stellar explosions of SNe~Ia make them useful as standardizable candles in measuring extragalactic distances \citep{phi93,riess96,wxf05,guy05,howell2011,burns18}, leading to the discovery of the accelerating of the Universe \citep{riess98,per99}. In recent years, larger samples of SNe~Ia have been used to further constrain the nature of dark energy driving the acceleration  \citep[e.g.,][]{betoule14,abbott19}.

However, the progenitor systems and explosion mechanism of SNe~Ia still remain controversial \citep[e.g.,][]{maoz14}. Two popular scenarios are the violent merger-triggered explosion of two WDs, known as the double-degenerate (DD) scenario \citep{DD1,DD2}, and the accretion-triggered explosion of a WD with a nondegenerate companion, known as the single-degenerate (SD) scenario \citep{SD2,nom82,nom97}. In general, the detection of signatures of circumstellar material (CSM) around some SNe Ia supports the SD scenario \citep{ham03,alder06,pat07,ster11,dil12,magu13,silver13,wxf2019}, though some theoretical studies show that the CSM could be also produced in the DD scenario \citep{shen13,raskin2013}. On the other hand, there is also evidence for nondetections of companion signatures for some SNe~Ia, thus favoring the DD scenario \citep{Gonz2012,Schaefer12,Olling15,tucker19}.

Popular explosion models of SNe~Ia include the following cases.
(1) The CO WD accretes material from the companion star until its mass nearly reaches the Chandrasekhar mass limit ($M_{\rm Ch}$, $\sim$1.4\,\Msun, \citealp{chand57}) and compressional heating at the center causes the explosion \citep{pier04}.
(2) The detonation of a thin layer of He on the surface of a WD \citep{kromer10,shen14} triggers a second detonation in the WD center and hence the explosion of a sub-$M_{\rm Ch}$ mass C-O WD.
(3) The violent merger or secular merger of two WDs, accompanied by radiation of gravitational waves \citep{Ropke12,gar17}.
(4) In triple systems, oscillations of the third star cause a direct collision of two WDs and trigger the SN explosion \citep{thom11,mazz18}.
In view of these explosion mechanisms, the delayed detonation model is one of the most suitable ones to account for the observed properties of SNe~Ia, which initially involves a deflagration of a $M_{\rm Ch}$ CO WD and later a supersonic detonation \citep{khok91,hoef17}. Nevertheless, the double detonation model of sub-$M_{\rm Ch}$ CO WDs shows many striking features, and can also explain the observed properties of some SNe~Ia \citep{shen18}.

Observationally, there is increasing evidence for spectroscopic and photometric diversity of SNe~Ia.
Most SNe~Ia can be classified as spectroscopically normal ones, while a small fraction exhibit peculiar properties in some respects \citep[e.g.,][]{bran93,fili97}, such as the SN~1991T-like overluminous SNe \citep{fili92a,rui92,phi93}, the SN~1991bg-like subluminous SNe \citep{fili92b,lei93}, or the SN~2002cx-like subclasses \citep{fili03,LiWD03}. Based on differences in \SiII\ velocity evolution, \citet{ben05} divided SNe~Ia into three subclasses: high velocity gradients (HVG), low velocity gradients (LVG), and FAINT. According to the equivalent width (EW) of \SiII~$\lambda$6355 and \SiII~$\lambda$5972 absorption lines near maximum brightness, \citet{bran06} divided SNe~Ia into core normal (CN), broad line (BL), cool (CL), and shallow silicon (SS) subgroups. The sublumious SN~1991bg-like and overluminous SN~1991T-like SNe~Ia have large overlap with the CL and SS subclasses, respectively \citep{bran06}. Based on the \SiII~$\lambda$6355 velocity measured near the time of $B$-band maximum, \citet{wxf09b} classified SNe~Ia into normal-velocity (NV) and high-velocity (HV) subsets. The HV subclass is found to share some common properties such as red $B-V$ color, slow decay in blue bands starting at $t \approx 40$\,d from the peak, and abundant surrounding CSM \citep{wxf08,wxf09a,wxf2019,fol11,mand14}. Although asymmetric explosions have been proposed to explain the occurrence of HV and NV subclasses of SNe~Ia \citep{Maeda10}, it is difficult to account for the fact that these two subgroups have different birth environments \citep{wxf13}.

Early-time observations can place important constraints on the explosion physics of SNe~Ia, including the size of the primary WD \citep{bloom12}, the radius of the companion star \citep{hoss17}, the distribution of \Nifs\ in the ejecta, and the possible existence of CSM \citep{piro16}. Therefore, clarifying the progenitor systems and explosion mechanisms affects our understanding of stellar evolution and precision cosmology. The unburned carbon detected in early-time spectra can provide important clues to the progenitor system and explosion mechanism of SNe~Ia \citep{Yamanaka2009,sliver11,taub11,thomas11,silver12a,hsiao13,liwx19}.

Previous studies show that nearly 30\% of SNe~Ia show signatures of \CII~$\lambda$6580 absorption at $t \approx -4$\,d (relative to the time of maximum light), while this fraction is over 40\% when examining the $t \approx -10$\,d spectra \citep{parr11,thomas11,fola12,silver12a,magu14}. These studies show that carbon-positive SNe~Ia tend to be LVG subtypes \citep{fola12} and have bluer optical colors around maximum light \citep{thomas11,silver12a}. Among those carbon-positive SNe~Ia, there are two events which show carbon absorption lasting until 1--3 weeks after maximum light. One is SN~2002fk, which has detecable carbon absorption lines in the $t \approx 10$\,d spectrum \citet{cart14}. Another example is SN 2018oh studied by \citet{liwx19}, the first {\it Kepler}-discovered SN~Ia with a spectroscopic classification; the carbon feature can be detected even in the $t \approx 20.5$\,d spectrum, representing the latest detection of carbon in SNe~Ia. The origin of these carbon detections in post-maximum spectra still remains unclear. SN~2017hpa is the third SN~Ia with persistent carbon feature; it exploded in the spiral galaxy UGC 3122 (see Fig. 1) at a distance of $\sim$~65.6\,Mpc (redshift $z \approx 0.0156$). The prominent carbon features and small distance of SN 2017hpa provide us with another excellent chance to study the observed diversity of SNe~Ia.

In this paper, the optical observations and data reduction are presented in Section 2. Section 3 discusses the light and color curves, while Section 4 shows the spectroscopic evolution. The quasibolometric light curve and origin of prominent carbon feature of SN 2017hpa are discussed in Section 5. We summarize in Section 6.

\section{Observations and Data Reduction}

\subsection{Discovery and Host Galaxy}

\begin{figure*}[ht]
\centering
\includegraphics[angle=0,width=172mm]{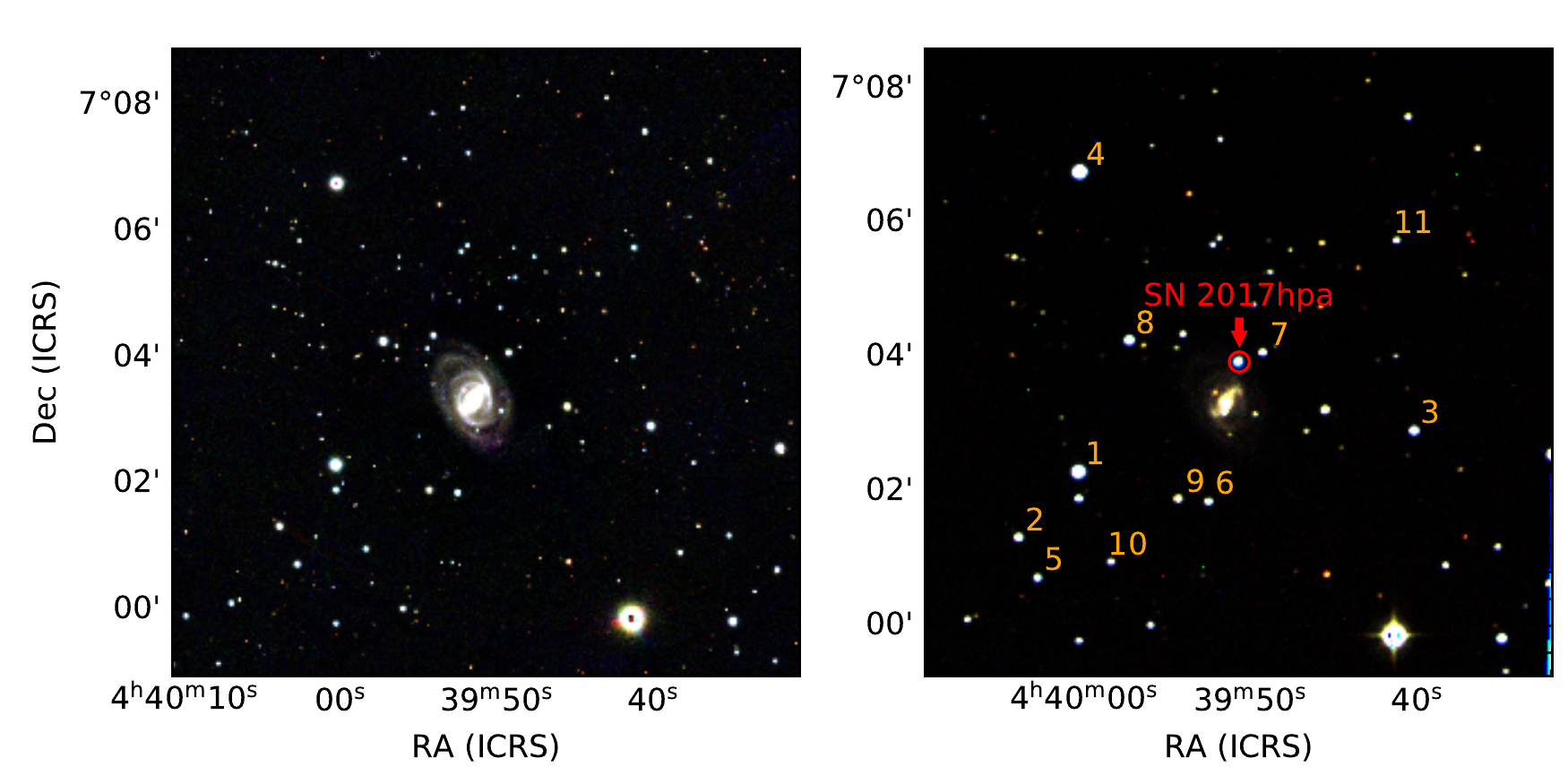}
\caption{The left panel shows a color image synthesized from $gri$-band observations from PanSTARRS, and the faint star right beside the position of SN 2017hpa is totally covered in following observations. The right panel shows a color image synthesized from $gri$-band observations from TNT; SN 2017hpa is marked with a red circle while the reference stars are numbered. \label{fobsimg}}
\end{figure*}

SN 2017hpa was discovered at $\alpha =04^h39^m50^s.750$, $\delta = 07^{\circ}03\arcmin54\arcsec.90$ (J2000) on 2017 Oct. 25.35 (UT dates are adopted throughout this paper) during the Puckett Observatory World Supernova Search (POSS) \citet{gag17}. Figure \ref{fobsimg} shows a color image of SN 2017hpa. A spectrum taken $\sim 0.65$\,d after the discovery classified it as a normal SN~Ia \citep{floers17}. 

The host galaxy of SN 2017hpa is UGC 3122, which is classified as SAB(rc)s at $z = 0.015631 \pm 0.000005$ \citep{pat02,spri05}. This redshift corresponds to a distance modulus $\mu = 34.05 \pm 0.38$\,mag with a velocity uncertainty of 500~\kms\ \citep{will97}, assuming a Hubble constant of 73.5~\kms\,Mpc$^{-1}$ \citep{riess18}.

\subsection{Photometry}
 
After the discovery of SN 2017hpa, we triggered follow-up photometric observations on several telescopes, including the 0.8\,m  Tsinghua-NAOC telescope (TNT; \citealp{huang12,zhang15}), the Las Cumbres Obervatory (LCO) Telescope network \citep{Shporer2011,brown13}, the 0.76\,m Katzman Automatic Imaging Telescope (KAIT) at Lick Observatory \citep{LiWD01,fili01,fili05}, and the 1\,m Nickel reflector at Lick Observatory. The TNT, Nickel, and KAIT monitored SN~2017hpa in the $BVRI$ bands, while the LCO 1\,m telescope sampled its light curves in the $BVgri$ bands.

For photometric observations obtained from the LCO during the Global Supernova Project, PyZOGY \citep{zackay16,guevel17} is employed for image subtraction while the $lcogtsnpipe$ \citep{valenti16} is applied for measuring the SN flux. An {\it ad hoc} pipeline (based on the IRAF DAOPHOT package; \citealp{stet87}) is applied to reduce images from the TNT and extract instrumental magnitudes of the SN. Photometric images from the Lick Observatory are reduced using the Lick Observatory Supernova Search (LOSS; \citealp{fili01}) data-reduction pipeline \citep{gane10,Stahl2019,Stahl2020phot}, while DAOPHOT \citep{stet87} is applied to implement the point-spread-function (PSF) photometry.

For the TNT instrumental magnitudes, the $BV$-band images are calibrated using the APASS catalog \citep{henden16} and the $gri$-band magnitudes are calibrated using the PanSTARRS catalog \citep{panstarr1,panstarr2,panstarr3,panstarr4}. The local standard stars with photometric magnitudes from APASS and PanSTARRS are listed in Table \ref{tab:standards}. The unfiltered instrumental magnitudes from LOSS are calibrated to the standard Landolt $R$-band magnitudes based on the transformed local standards of SDSS \citep{LiWD03,zheng17b}. The LOSS $BVRI$ instrumental magnitudes are calibrated to the Johnson system using a series of Landolt \citep{land92} standard stars taken on a number of photometric nights.

\begin{deluxetable*}{lrrrrrrr}
\tablecaption{Photometric Standards in the SN 2017hpa Field 1\tablenotemark{a}}\label{tab:standards}
\tablehead{\colhead{Star} &\colhead{$\alpha$(J2000)} &\colhead{$\delta$(J2000)} &\colhead{$B$ (mag)} &\colhead{$V$ (mag)} &\colhead{$g$ (mag)} &\colhead{$r$ (mag)} &\colhead{$i$ (mag)} }
\startdata
1  & 04:40:00.331 &+07:02:17.167 &14.125(020) &13.368(015) &13.689(014) &13.114(034) &12.915(040) \\
2  & 04:40:03.904 &+07:01:18.451 &16.618(065) &15.697(037) &16.160(089) &15.398(107) &15.110(070) \\
3  & 04:39:40.185 &+07:02:54.100 &15.836(029) &15.046(049) &15.403(030) &14.779(034) &14.593(045) \\
4  & 04:40:00.268 &+07:06:44.968 &14.140(034) &13.228(016) &13.625(016) &12.887(036) &12.632(035) \\
5  & 04:40:02.771 &+07:00:42.491 &16.812(121) &16.013(077) &16.389(088) &15.752(053) &15.491(174) \\
6  & 04:39:52.514 &+07:01:50.567 &17.515(169) &16.242(070) &16.645(040) &15.957(026) &15.637(040) \\
7  & 04:39:49.255 &+07:04:03.612 &17.125(095) &16.130(077) &16.594(044) &15.802(069) &15.477(068) \\
8  & 04:39:57.276 &+07:04:14.560 &15.626(052) &14.823(023) &15.191(018) &14.556(044) &14.326(043) \\
9  & 04:39:54.303 &+07:01:52.794 &17.553(151) &16.262(008) &16.918(114) &15.686(037) &15.101(034) \\
10 & 04:39:58.351 &+07:00:56.540 &17.480(238) &16.547(158) &16.996(066) &16.230(041) &15.995(040) \\
11 & 04:39:41.224 &+07:05:43.976 &17.430(026) &16.468(078) &17.033(093) &16.288(066) &16.102(040) \\
\enddata
\tablenotetext{a}{Standard stars used for calibration of instrumental magnitudes.}
\end{deluxetable*}

\begin{figure}[ht]
\centering
\includegraphics[angle=0,width=86mm]{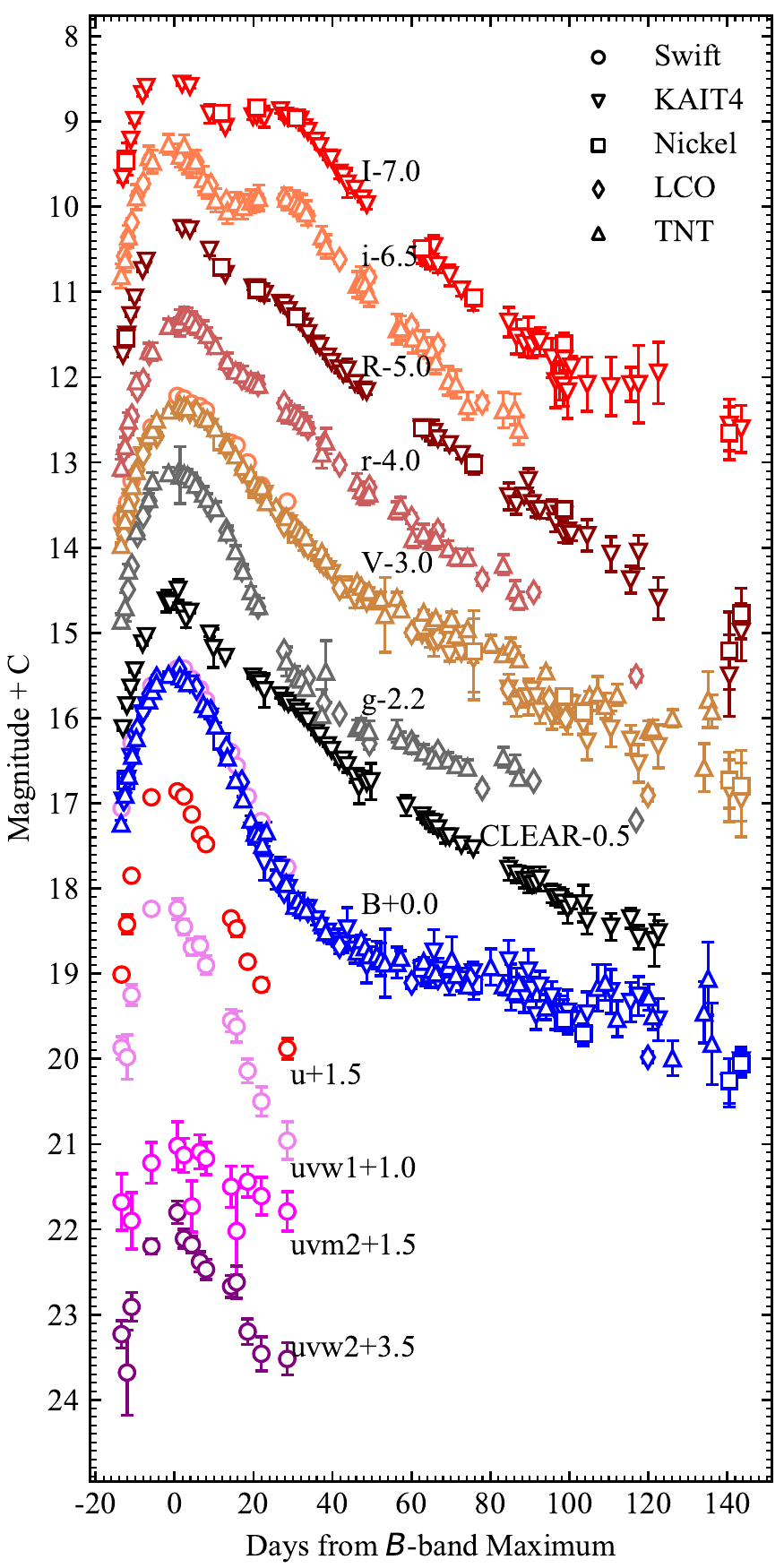}
\caption{The observed UV and optical light curves of SN 2017hpa. \label{fmultlcv}}
\end{figure}

Optical and ultraviolet (UV) observations of SN 2017hpa were also obtained with the Neil Gehrels {\it Swift} Observatory \citep{geh04}. The {\it Swift}/UVOT observations started at relatively early phases in six bands including $uvw2$, $uvm2$, $uvw1$, $u$, $b$, and $v$ \citep{rom05}. The filters in lower case are used throughout this paper for photometric observations in UVOT bands. Using zeropoints extracted by \citet{bree11} in the Vega system, the data-reduction pipeline of the {\it Swift} Optical/Ultraviolet Supernova Archive (SOUSA; \citealp{brown14}) is applied to obtain the {\it Swift} optical/UV light curves of SN~2017hpa. The source counts are measured using a 3$\arcsec$ aperture and corrections are based on the average PSF. The template-subtraction technique has also been applied to the {\it Swift} images and the final uncertainty in the photometry is the combination of statistical uncertainties in galaxy subtraction count rates and a 2\% systematic fluctuation at each pixel caused by differences in response sensitivity across the photon detector. The final observed {\it Swift} and ground-based light curves are shown in Figure \ref{fmultlcv}, and the corresponding magnitudes are tabulated in Table \ref{tab:gndphot} and Table \ref{tab:swiftphot}.

\startlongtable
\begin{deluxetable*}{lccccccccccccc}
\tablecolumns{0} \tablewidth{0pc} \tabletypesize{\scriptsize}
\tablecaption{Photometric Observations of SN 2017hpa by Ground-Based Telescopes \label{tab:gndphot}}
\tablehead{\colhead{MJD} &\colhead{Epoch\tablenotemark{a}} &\colhead{$B$ (mag)} &\colhead{$V$ (mag)} &\colhead{$R$ (mag)} &\colhead{$I$ (mag)} &\colhead{$g$ (mag)} &\colhead{$r$ (mag)} &\colhead{$i$ (mag)} &\colhead{$Clear$ (mag)} &\colhead{Telescope}}
\startdata
58053.34  & -13.30    & 17.235(035) & 16.967(022) & \nodata  & \nodata  & 17.054(081) & 17.060(097) & 17.311(143) & \nodata  & TNT       \\
58053.54  & -13.10    & 16.971(031) & 16.869(027) & 16.741(023) & 16.668(039) & \nodata  & \nodata  & \nodata  & 16.623(022) & KAIT4      \\
58053.82  & -12.83    & 16.946(041) & 16.848(030) & \nodata  & \nodata  & 16.981(025) & 16.863(065) & 17.078(043) & \nodata  & LCO     \\
58054.35  & -12.29    & 16.897(033) & 16.684(020) & \nodata  & \nodata  & 16.896(167) & 16.806(102) & 17.122(136) & \nodata  & TNT       \\
58054.36  & -12.28    & 16.741(012) & 16.634(008) & 16.539(010) & 16.463(015) & \nodata  & \nodata  & \nodata  & \nodata  & Nickel    \\
58054.36  & -12.28    & 16.744(012) & 16.642(011) & 16.546(011) & 16.463(016) & \nodata  & \nodata  & \nodata  & \nodata  & KAIT4      \\
58054.54  & -12.11    & 16.717(027) & 16.614(021) & 16.505(019) & 16.442(190) & \nodata  & \nodata  & \nodata  & 16.355(024) & KAIT4      \\
58054.84  & -11.80    & 16.705(043) & 16.659(042) & \nodata  & \nodata  & 16.692(022) & 16.614(058) & 16.856(045) & \nodata  & LCO     \\
58055.19  & -11.45    & 16.674(031) & 16.484(017) & \nodata  & \nodata  & 16.469(074) & 16.507(089) & 16.846(130) & \nodata  & TNT       \\
58055.54  & -11.10    & 16.460(021) & 16.409(017) & 16.276(017) & 16.217(030) & \nodata  & \nodata  & \nodata  & 16.147(016) & KAIT4      \\
$\vdots$  & $\vdots$  & $\vdots$    & $\vdots$    & $\vdots$    & $\vdots$  & $\vdots$  & $\vdots$  & $\vdots$  & $\vdots$  & $\vdots$       \\
58203.00  & +136.35   & 19.821(478) & 18.920(192) & \nodata  & \nodata  & \nodata  & \nodata  & \nodata  & \nodata  & TNT       \\
58207.17  & +140.52   & 20.258(264) & 19.730(330) & 20.210(455) & 19.657(309) & \nodata  & \nodata  & \nodata  & \nodata  & Nickel    \\
58207.17  & +140.52   & 20.281(285) & 19.851(363) & 20.501(477) & 19.562(302) & \nodata  & \nodata  & \nodata  & \nodata  & KAIT4      \\
58210.16  & +143.52   & 20.056(159) & 19.796(416) & 19.774(295) & \nodata  & \nodata  & \nodata  & \nodata  & \nodata  & Nickel    \\
58210.16  & +143.52   & 20.021(159) & 19.960(435) & 19.991(337) & 19.606(277) & \nodata  & \nodata  & \nodata  & \nodata  & KAIT4      \\
\enddata
\tablenotetext{a}{Relative to the epoch of $B$-band maximum brightness (MJD = 58,066.6). Magnitudes are calibrated to AB magnitude system.}
\end{deluxetable*}

\begin{deluxetable*}{lrrrrrrr}
\tablecaption{{\it Swift} UVOT Photometry of SN 2017hpa}\label{tab:swiftphot}
\tablehead{\colhead{MJD}& \colhead{Epoch\tablenotemark{a}} &\colhead{$uvw2$ (mag)} &\colhead{$uvm2$ (mag)} &\colhead{$uvw1$ (mag)} &\colhead{$u$ (mag)} &\colhead{$b$ (mag)} &\colhead{$v$ (mag)} }
\startdata
58053.19 & -13.45 & 19.73(16) & 20.18(33) & 18.87(13) & 17.51(06) & 17.06(04) & 16.66(05)  \\
58054.59 & -12.06 & 20.18(50) &  \nodata  & 18.98(26) & 16.92(11) & 16.73(07) & 16.47(11)  \\
58055.71 & -10.93 & 19.41(17) & 20.40(33) & 18.25(12) & 16.35(05) & 16.29(04) & 16.22(06)  \\
58060.76 &  -5.89 & 18.70(09) & 19.72(24) & 17.24(06) & 15.43(03) & 15.62(03) & 15.59(04)  \\
58067.35 &   0.70 & 18.30(13) & 19.52(28) & 17.24(12) & 15.36(05) & 15.42(04) & 15.22(06)  \\
58069.00 &   2.35 & 18.61(11) & 19.63(20) & 17.45(09) & 15.42(03) & 15.42(03) & 15.25(04)  \\
58071.00 &   4.34 & 18.68(10) & 20.23(30) & 17.69(10) & 15.63(04) & 15.54(03) & 15.31(04)  \\
58073.05 &   6.41 & 18.88(12) & 19.59(20) & 17.67(10) & 15.87(04) & 15.66(03) & 15.34(04)  \\
58074.59 &   7.95 & 18.97(12) & 19.67(19) & 17.90(11) & 15.98(04) & 15.79(03) & 15.39(04)  \\
58080.95 &  14.31 & 19.17(13) & 20.00(24) & 18.55(13) & 16.85(06) & 16.40(04) & 15.77(04)  \\
58082.35 &  15.71 & 19.12(19) & 20.52(51) & 18.62(18) & 16.97(09) & 16.56(05) & 15.80(07)  \\
58085.14 &  18.49 & 19.70(15) & 19.94(18) & 19.14(14) & 17.36(06) & 16.92(04) & 16.00(04)  \\
58088.60 &  21.96 & 19.96(20) & 20.11(22) & 19.50(17) & 17.63(08) & 17.21(05) & 16.27(05)  \\
58095.17 &  28.52 & 20.02(19) & 20.29(23) & 19.96(22) & 18.38(12) & 17.75(06) & 16.46(06)  \\
\enddata
\tablenotetext{a}{Relative to the epoch of $B$-band maximum (MJD = 58,066.6). Magnitudes are calibrated to Vega magnitude system.}
\end{deluxetable*}

\subsection{Spectroscopy}

A total of 26 low-resolution optical spectra of SN 2017hpa have been obtained using different telescopes and equipment, including the AFOSC mounted on the Asiago Ekar telescope, the BFOSC mounted on the Xinglong 2.16\,m telescope (XLT; \citealp{Jiang99,XLT2015,FanZh16}), the YFOSC on the Lijiang 2.4\,m telescope (LJT; \citealp{Chen2001,ZhangW12,wcj19}) of Yunnan Astronomical Observatories, the Kast spectrograph on the Lick 3\,m Shane telescope \citep{miller93,Stahl2020phot}, and the LCO 2\,m Faulkes Telescope North (FTN; \citealp{brown13}). The journal of spectroscopic observations is presented in Table \ref{tab:spectra}, including one spectrum from the Transient Name Server\footnote{https://wis-tns.weizmann.ac.il/} and six spectra from \citet{Stahl2020phot}. When no data-reduction pipeline for was available, we applied standard IRAF routines to reduce the spectra. Spectrophotometric standard stars observed at an airmass comparable to the target on the same night were used to calibrate the flux density of SN 2017hpa. The extinction curves of the various observatories are utilized to correct for atmospheric extinction, and spectra of the standard stars are used to eliminate the telluric absorption lines.

\begin{deluxetable*}{lrrrr}
\tablecaption{Spectroscopic Observations of SN 2017hpa}\label{tab:spectra}
\tablehead{\colhead{MJD}& \colhead{Epoch\tablenotemark{a}} &\colhead{$\lambda_{\rm Start}$} &\colhead{$\lambda_{\rm End}$} &\colhead{Instrument} }
\startdata
58052.5 & -14.1 & 3387 &  8249 & Asiago (public)  \\
58053.2 & -13.5 & 3496 &  9173 & LJT     \\
58054.2 & -12.4 & 3502 &  9171 & LJT     \\
58056.5 & -10.1 & 3622 & 10400 & Lick 3\,m  \\
58058.7 &  -7.9 & 3399 &  9999 & LCO     \\
58064.5 &  -2.1 & 3249 &  9999 & LCO     \\
58068.7 &   2.0 & 3249 &  9999 & LCO     \\
58071.3 &   4.7 & 3746 &  8840 & XLT     \\
58075.4 &   8.7 & 3250 & 10000 & LCO     \\
58076.1 &   9.5 & 3744 &  8839 & XLT     \\
58078.4 &  11.7 & 3630 & 10400 & Lick 3\,m  \\
58091.7 &  25.1 & 3500 &  9100 & LJT     \\
58094.3 &  27.7 & 3299 &  9999 & LCO     \\
58099.3 &  32.6 & 3299 &  9999 & LCO     \\
58099.3 &  32.6 & 3630 & 10400 & Lick 3\,m  \\
58105.4 &  38.7 & 3632 & 10400 & Lick 3\,m  \\
58107.7 &  41.1 & 3500 &  9100 & LJT      \\
58110.3 &  43.7 & 5896 &  8182 & Lick 3\,m  \\
58111.3 &  44.7 & 3249 & 10000 & LCO     \\
58118.7 &  52.1 & 3500 &  9100 & LJT     \\
58119.4 &  52.8 & 3300 & 10000 & LCO     \\
58121.3 &  54.7 & 3299 & 10000 & LCO     \\
58127.8 &  61.1 & 3500 &  9100 & LJT    \\
58131.4 &  64.7 & 3632 & 10400 & Lick 3\,m  \\
58132.3 &  65.7 & 3400 &  9300 & LCO     \\
58153.3 &  86.6 & 3300 &  9299 & LCO     \\
\enddata
\tablenotetext{a}{Relative to the epoch of $B$-band maximum (MJD = 58,066.6).}
\end{deluxetable*}

\section{Light Curves}

\subsection{Optical and Ultraviolet Light Curves}

\begin{figure*}[ht]
\centering
\includegraphics[angle=0,width=140mm]{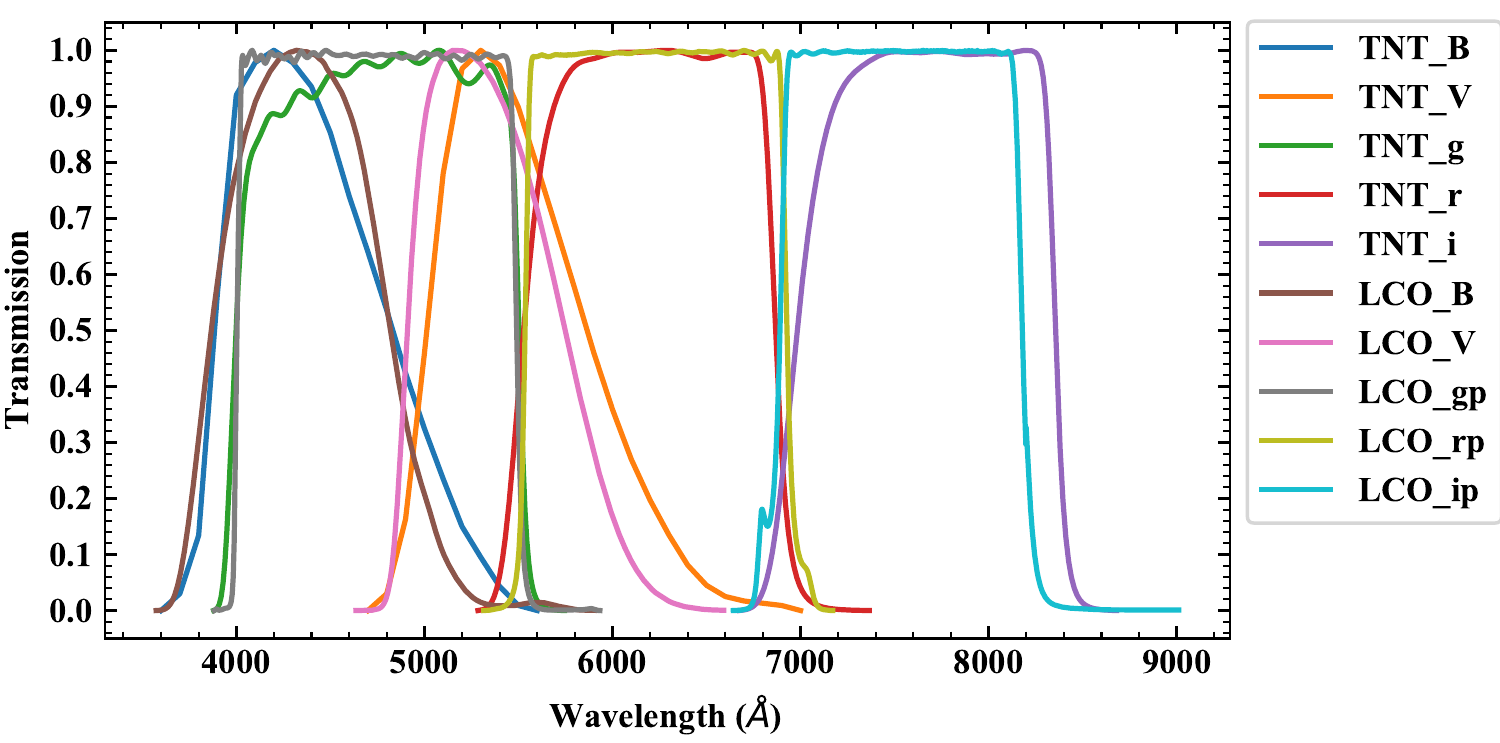}
\caption{The transmission curves of the TNT and LCO filters; each curve is normalized to the peak. \label{filtercurves}}
\end{figure*}

The multiband UV/optical light curves of SN 2017hpa are shown in Figure \ref{fmultlcv}; one can see that the observations in optical bands have nearly daily sampling, ranging from about 2 weeks before to over 100\,d after $B$-band maximum light. The light curves of SN 2017hpa are similar to those of normal SNe~Ia,
reaching maximum slightly earlier in the $I/i$ and $UV$ bands than the $B$ band, and having a prominent shoulder in $R/r$ as well as a secondary maximum in $I/i$. The slight deviations between the $BVgri$ light curves of different telescopes are primarily due to different filter transmission functions, as shown in Figure \ref{filtercurves}. The transmission differences at the red edge of the $I$-band filters may cause the $I$-band discrepancies between LCO and TNT. Applying a polynomial fit to the $B$-band light curves around maximum light yields a peak of $15.48 \pm 0.03$\,mag on MJD = 58066.6 (UT 2017 November 9.6). The $V$-band light curve reached its peak of $15.35 \pm 0.2$\,mag on MJD = 58068.4, $\sim 1.8$\,d after the $B$-band peak.

\begin{figure*}[ht]
\centering
\includegraphics[angle=0,width=172mm]{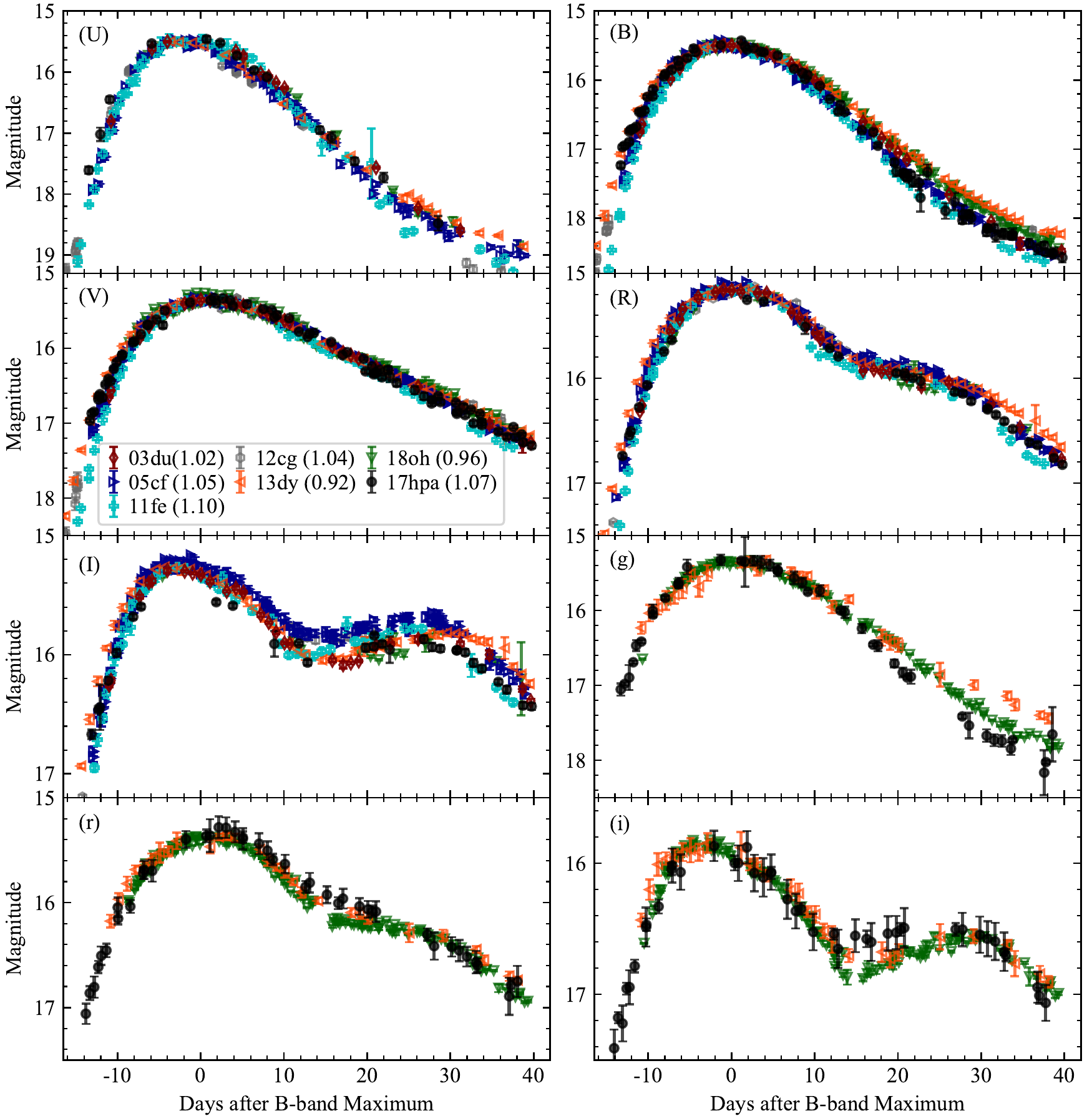}
\caption{Comparison of the optical light curves of SN 2017hpa with other well-observed SNe~Ia having similar decline rates. The light curves of the comparison SNe~Ia have been normalized to match the observed peak magnitudes of SN 2017hpa. \label{foptcomp}}
\end{figure*}

\begin{figure*}[ht]
\centering
\includegraphics[angle=0,width=172mm]{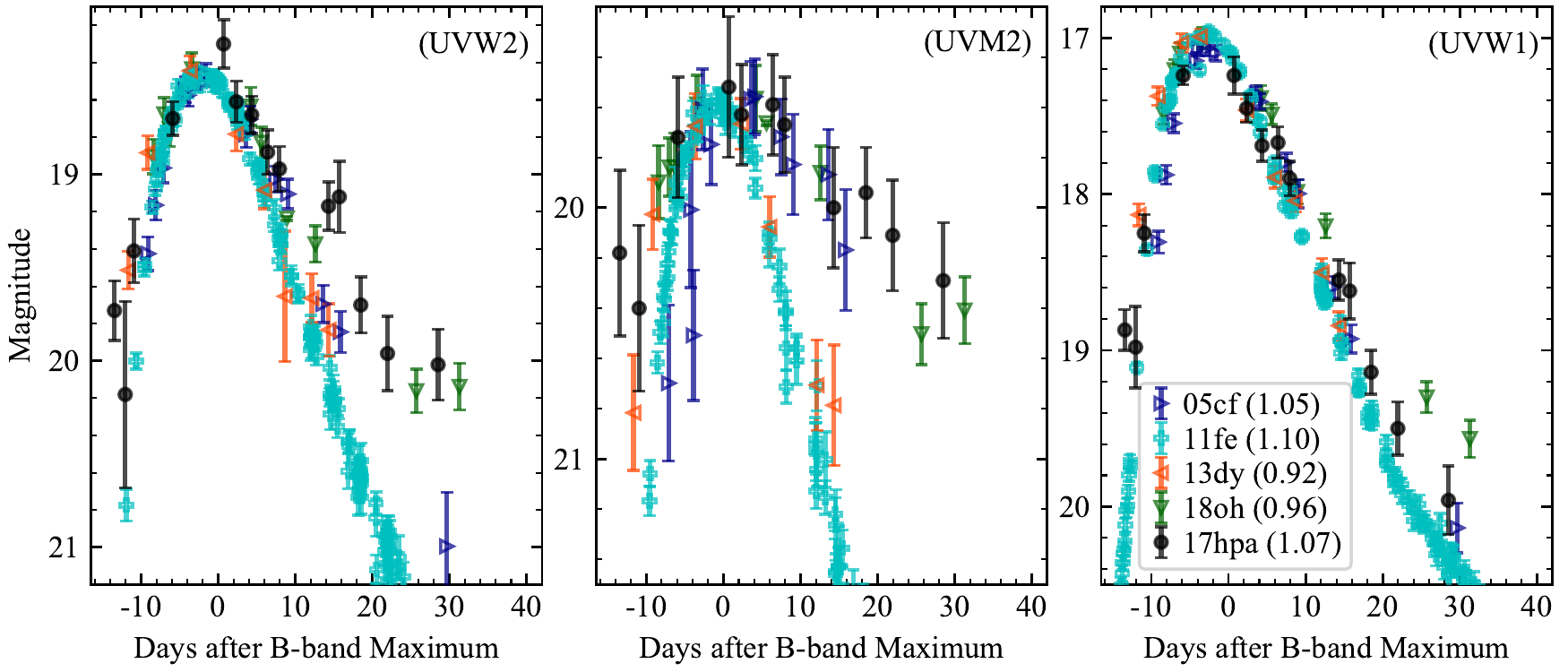}
\caption{Comparison of the UV light curves of SN 2017hpa with other well-observed SNe~Ia having similar decline rates. The light curves of the comparison SNe~Ia have been normalized to match the peak magnitudes of SN 2017hpa. \label{fuvcomp}}
\end{figure*}

Figures \ref{foptcomp} and \ref{fuvcomp} compare the multiband UV/optical light curves of SN 2017hpa with those of several well-observed normal SNe~Ia which have comparable \mb, including SN 2003du\citep{stan07}, SN 2005cf \citep{wxf09b}, SN 2011fe \citep{magu13}, SN 2012cg \citep{munari13,brown14}, SN 2013dy \citep{pan15,zhai16}, and SN 2018oh \citep{liwx19}. The UV/optical light curves of the comparison SNe~Ia have been normalized to SN 2017hpa. As can be seen from Figure \ref{foptcomp}, SN 2017hpa and other normal comparision SNe~Ia have similar light-curve shapes near $B$-band maximum. Although the UV light curves of SN 2017hpa are similar to those of other comparison SNe~Ia, they seem to show excess emission at early phases, especially the first two data points. This may suggest additional energy beyond the radioactive decay of centrally-located nickel, such as surface nickel mixing \citep{Piro13} or interaction of SN ejecta with a companion star or with CSM \citep{Kasen10}. The post-peak decline rate \mb\ of the $B$-band light curve is measured to be \dmvalue\,mag, and the color stretch \citep{burns14} is determined to be \sBV = \sbvalue.

\subsection{Reddening and Color Curves}

Assuming $R_{V} = 3.1$ \citep{Cardelli89}, we obtain the line-of-sight Galactic extinction for SN 2017hpa to be $A_{V} = 0.485$\,mag \citep{sch98,sf11}, corresponding to a color excess of $E(B-V)_{\rm gal} = 0.156$\,mag. After removing the Galactic reddening, the $B-V$ color is found to be $0.002 \pm 0.05$\,mag at $t = 0$\,d and $1.08 \pm 0.06$\,mag at $t = 35$\,d relative to $B$ maximum, consistent with typical values of normal SNe~Ia \citep{phi99,wxf09b}.

\begin{figure}[ht]
\centering
\includegraphics[angle=0,width=86mm]{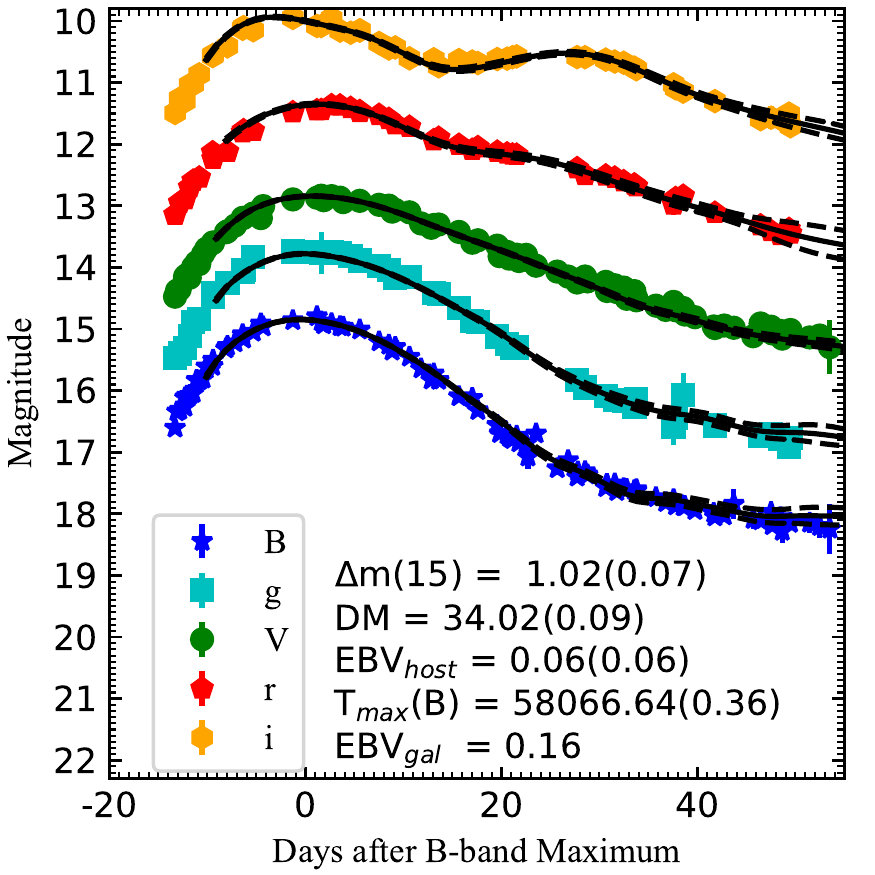}
\caption{Best-fit light-curve model from SNooPy2. The light curves are shifted vertically for clarity. The dashed lines represent the 1$\sigma$ uncertainty of the best-fit light-curve templates. \label{fsnpymodel}}
\end{figure}

We applied SuperNovae in object-oriented Python (SNooPy; \citealp{burns11,burns14}) to fit the multiband light curves of SN 2017hpa, as shown in Figure \ref{fsnpymodel}. Both $EBV$ and $st$ models in SNooPy2 are adopted to estimate the host-galaxy extinction, and an average host reddening is derived to be $E(B-V)_{\rm host}$ = \ebvalue\,mag. The relatively low host-galaxy reddening is consistent with the fact that the SN is located far away from the center of the host galaxy. Moreover, the spectra of SN 2017hpa show no detectable absorption feature of \NaI~D due to host galaxy.

\begin{figure*}[ht]
\centering
\includegraphics[angle=0,width=172mm]{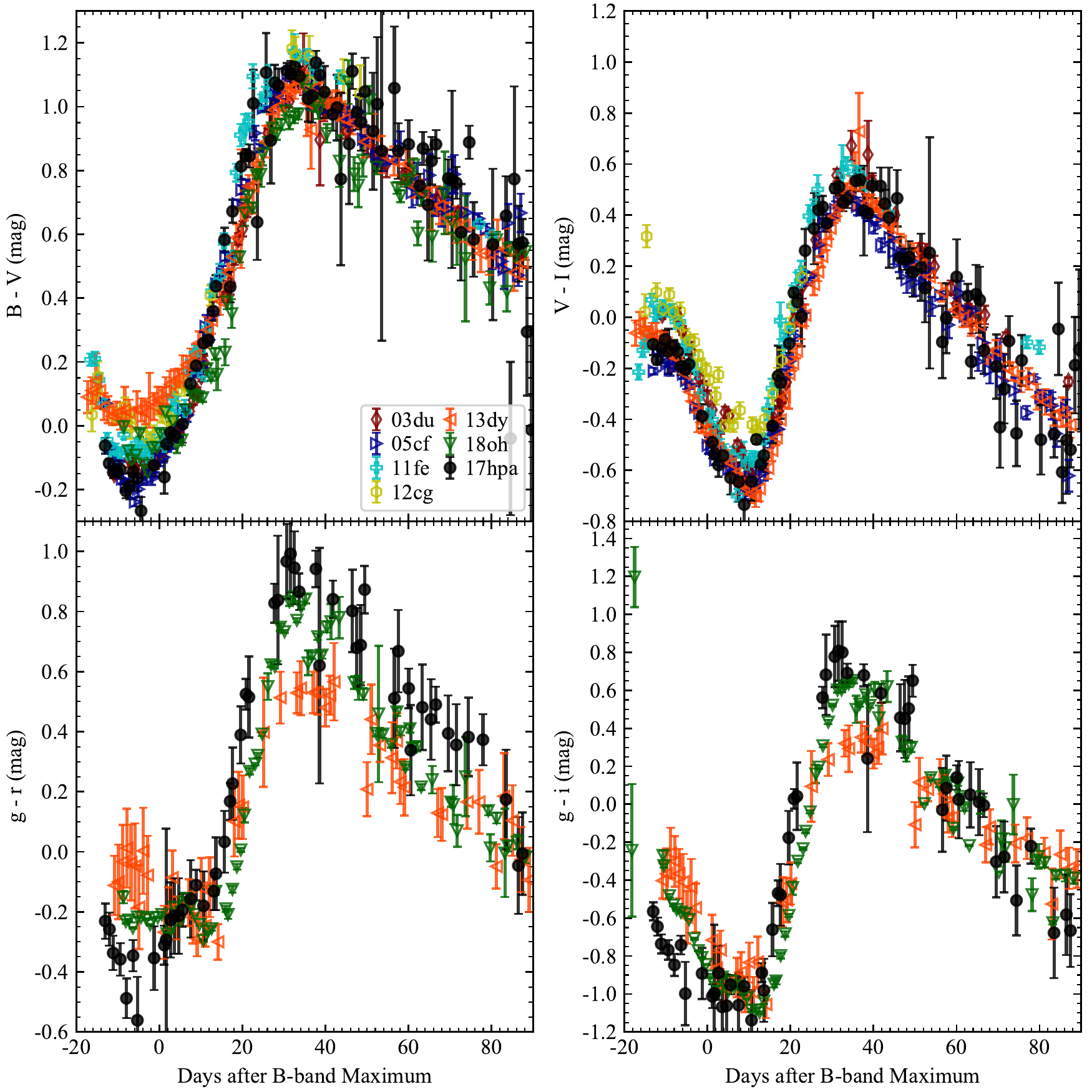}
\caption{The $B-V$, $V-I$, $g-r$, and $g-i$ color curves of SN 2017hpa compared with those of SNe 2003du, 2005cf, 2011fe, 2012cg, 2013dy, and 2018oh. All light curves including those of SN 2017hpa have been dereddened using SNooPy2. \label{fcolorcomp}}
\end{figure*}

The optical intrinsic color evolution of SN 2017hpa is shown in Figure \ref{fcolorcomp}.
At $t \gtrsim -10$\,d, both the $B-V$ and $g-r$ color curves evolve toward the red until reaching the reddest color at 4--5 weeks after $B$ maximum. Both the $V-I$ and $g-i$ color curves show a short-term evolution from red to blue until $t \approx -10$\,d; then they evolve redward and reach the red peak at $t \approx 35$\,d. After that, the $V-I$ and $g-i$ color curves became progressively bluer. 

\begin{figure}[ht]
\centering
\includegraphics[angle=0,width=86mm]{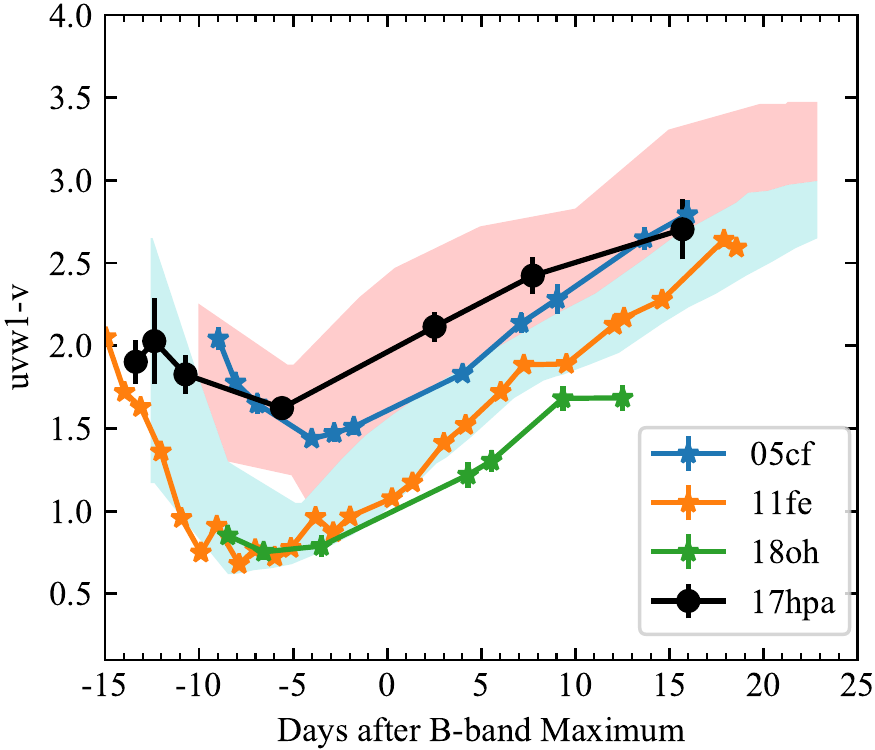}
\caption{The $uvw1-v$ color of SN 2017hpa compared to the group of NUV-blue and NUV-red SNe. The pink shaded region represents the regions covered by SNe~Ia that are classified as NUV-red and the blue shaded region represents the regions covered by SNe~Ia that are classified as NUV-blue. The overplotted curve is the unreddened color evolution of SN 2017hpa (see \citealp{mil13}). \label{fnuvblue}}
\end{figure}

Overall, the color-curve evolution of SN 2017hpa is similar to that of SN 2005cf and SN 2018oh, except it has a bluer color at very early phases (especially the $g-r$ color). Based on the near-UV (NUV) colors, SNe~Ia can be classified into NUV-red and NUV-blue subgroups \citep{mil13}. Figure \ref{fnuvblue} shows the observed $uvw1-V$ color evolution of SN 2017hpa together with that of SNe 2005cf, 2011fe, and 2018oh. One can see that SN 2017hpa can be put into the NUV-red group. 

\subsection{First-Light Time}

\begin{figure}[ht]
\centering
\includegraphics[angle=0,width=86mm]{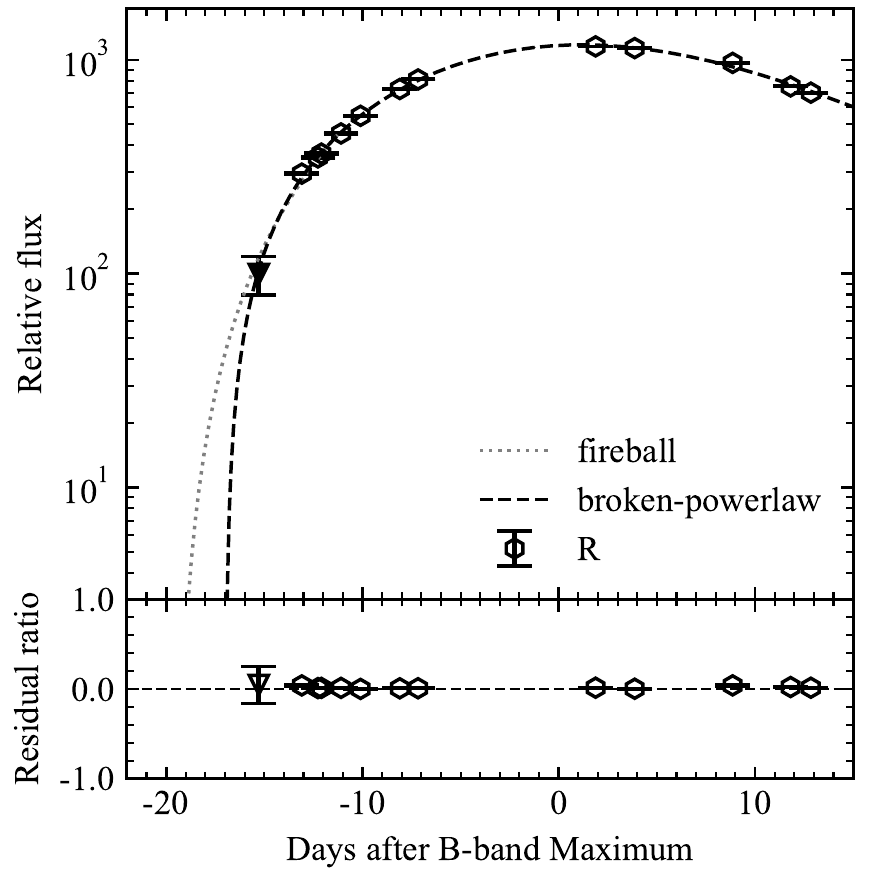}
\caption{Fit to the observed $R$-band light curves using the analytic function from \citet{zheng17a} and the ideal fireball model \citep{Riess1999}. The black triangle represents the discovery magnitude, i.e., 17.9 mag $\pm$ 0.3 mag in clear band (close to broadband R), obtained at $\sim$2.0 days before our multi-color observations. The bottom panel shows the residual relative to the best-fit curves. The horizontal dashed line in the bottom panel represents zero residual. \label{ftrise}}
\end{figure}

The rise time and first-light time can put additional constraints on the radius of the exploding star itself \citep{bloom12,Piro13}.
The observation on Oct. 13, 2017 by \citet{gag17} provides a non-detection limit magnitude of 20.5 mag. However, this observation is about 12 days before the discovery and can not provide very useful constraint for the explosion date and hence the rise time of the light curves. We thus only utilized the discovery magnitude, i.e., 17.9 mag $\pm$ 0.3 mag in clear band (close to broadband $R$), obtained at $\sim$2.0 days before our multi-color observations when performing the rise time fitting. The ideal expanding fireball \citep{Riess1999} model and broken-power-law \citep{zheng17a} model are both adopted to fit the $R$-band light curve of SN 2017hpa (as shown in Figure \ref{ftrise}), and the first light time of the light curve is estimated as 58047.08 $\pm$ 0.73 days and 58049.65 $\pm$ 0.24 days, respectively. The mean fitted first-light time (FFLT) is adopted as 58048.37 $\pm$ 0.97 days.

With the time of maximum light in $B$ and the derived FFLT, the rise time of SN 2017hpa is estimated to be \trise\,d, comparable to that of typical SNe~Ia \citep{zheng17b}. The first multi-color observation of SN 2017hpa is thus estimated to be $\sim 5$\,d after the FFLT, $\sim 13$\,d prior to $B$ maximum.

\section{Optical Spectra}

\subsection{Temporal Evolution of the Spectra}

\begin{figure}[ht]
\centering
\includegraphics[angle=0,width=86mm]{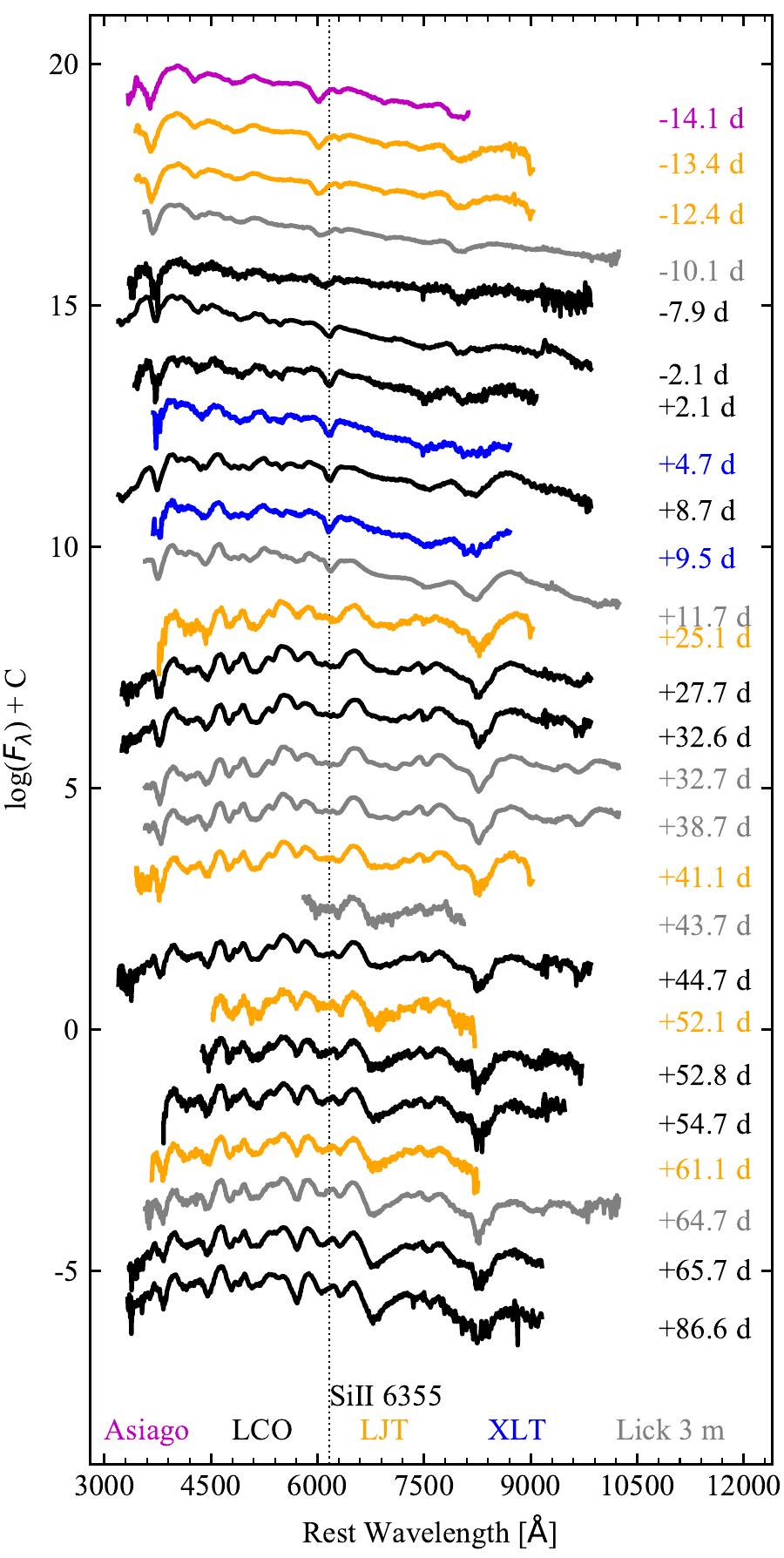}
\caption{Optical spectral evolution of SN 2017hpa. All of the spectra have been corrected for the redshift of the host galaxy and reddening. The epochs shown on the right side represent the phases in days relative to $B$-band maximum light. The dashed line marks the center of \SiII~$\lambda$6355 line profile at +2.04\,d from $B$-band maximum. The color of the spectrum stands for different instruments. The spectra have been shifted vertically for clarity. \label{fspecevolution}}
\end{figure}

\begin{figure*}[ht]
\centering
\includegraphics[angle=0,width=172mm]{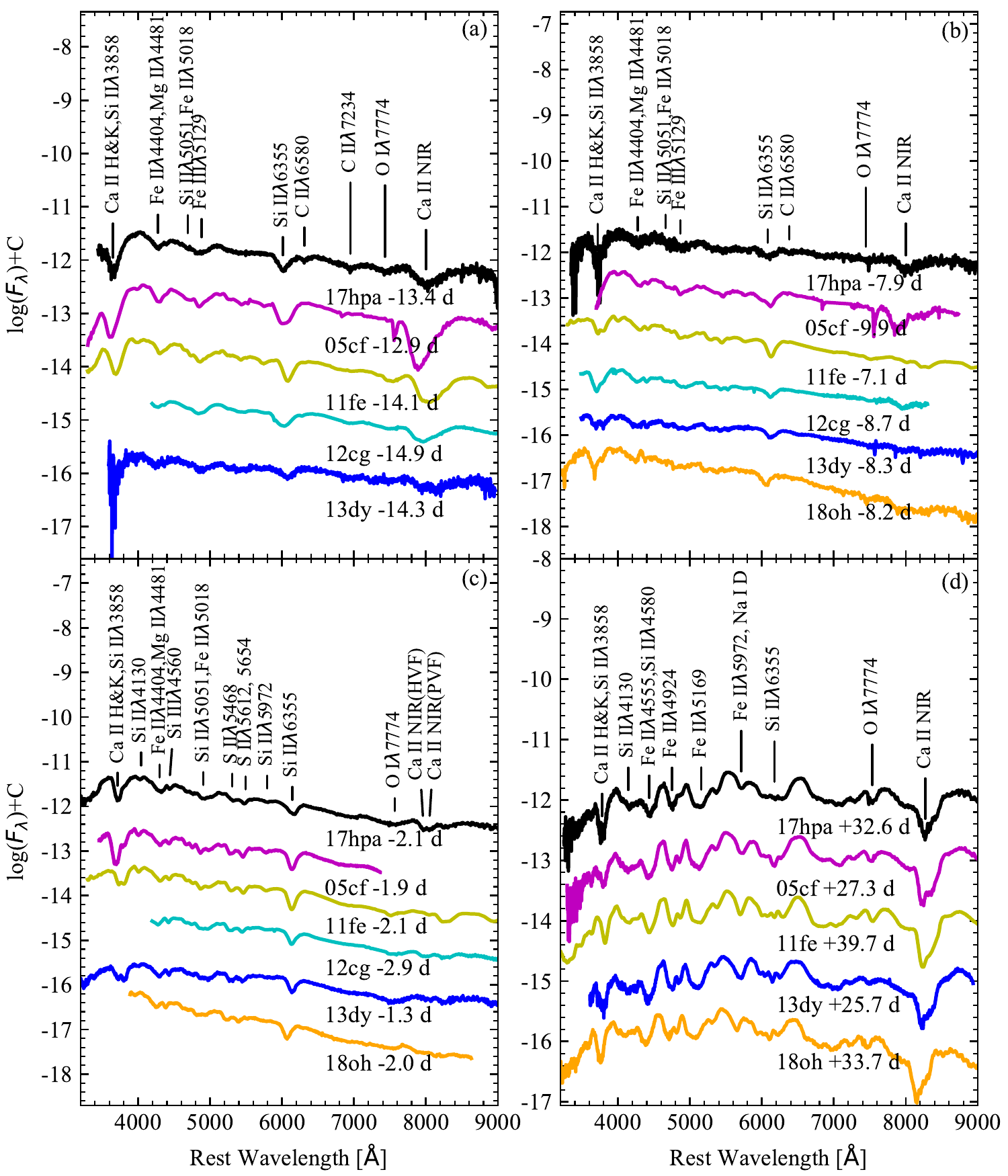}
\caption{Spectra of SN 2017hpa at $t \approx -14$, $-8$, -2, and +32\,d relative to $B$-band maximum, compared with spectra of SNe 2005cf \citep{gara07,wxf09a}, 2011fe \citep{mazz14,zhangkc16}, and 2013dy \citep{zheng13,pan15,zhai16} at comparable phases. Correction of reddening and redshift of the host galaxy had been done for all of the given spectra. The spectra have been shifted vertically for clarity.  \label{fspeccomp}}
\end{figure*}

The evolution of the optical spectra of SN 2017hpa is displayed in Figure \ref{fspecevolution}. 
The early-time spectra are characterized by prominent absorption lines of intermediate-mass elements (IMEs), such as \FeII\,$\lambda\lambda$4404,5018, \MgII\,$\lambda$4481, \SiII\,$\lambda$6355, \SII\,$\lambda\lambda$5468,5654, \CaII\,NIR triplet and \CaII\,H\&K.
At $t \sim$ 2 weeks before the $B$-band maximum, the absorption troughs near 4300 \Angst\ and 4800 \Angst\ could be attributed to \FeII/\FeIII/\MgII, while the distinct absorption notches near 6300 \Angst\ and 7000 \Angst\ could be due to \CII\,$\lambda$6580 and \CII\,$\lambda$7234, respectively. The \CII\,$\lambda$6580 absorption is relatively strong while the \CII\,$\lambda$7234 absorption is weaker. The \SiII\,$\lambda$6355 absorption lines at this phase display a perfect gaussian profile without invoking the high-velocity feature (HVF), while a \CaII\,NIR HVF could be detected through multi-gaussian fitting \citep{zhao15,zhao16}. After $t \sim$ 1 week before maximum light, both \CII\,$\lambda$6580 and \CII\,$\lambda$7234 absorptions are still prominent in the spectra of SN 2017hpa, and the absorption lines of ``W''-shaped \SII\ and \SiII\,$\lambda$5972 start to emerge in the spectra of SN 2017hpa. With the decreasing of the expansion velocity of the photosphere, the absorption minimum of \SiII\,$\lambda$6355 line gradually shifted redward, and the absorption lines of iron group elements and sulfur gradually increase in its strength.
At around $B$-band maximum, the spectra are primarily dominated by ``W''-shaped \SII\ absorption features near 5400\,\Angst, the blended absorption lines of \FeII\ and \SiII/\SiIII\ near 4500\,\Angst\ and \SiII\,$\lambda$6355, while the C II features become invisible at this phase. By $t \sim$ 0 days, the HVFs of \CaII\,NIR triplet become relatively weak and the photospheric component started to emerge in the spectra. At $t \sim +10$ days after the $B$ maximum, the photospheric components of the \CaII\,NIR continute to gain the strength and start to dominate the spectra features. Interestingly, the \CII\,$\lambda$6580 absorption feature seems to reemerge in the spectra of SN 2017hpa around this phase, which is rarely seen in other SNe Ia.
At about one month after the $B$-band maximum, the \CaII\,H\&K lines and NIR triplet are the main spectral features. Meanwhile, the features of iron group element begin to dominate in the spectra when the SN enter the early nebular phase.
Figure \ref{fspeccomp} compares spectra of SN 2017hpa at several epochs with those of well-observed SNe~Ia with similar \mb.

For SN 2017hpa, the \CII~$\lambda$6580 and \CII~$\lambda$7234 absorptions appear stronger than those of the comparison SNe~Ia in the early-phase spectra, as shown in Figure \ref{fspeccomp}(a). Moreover, the \OI\,$\lambda$7774 absorption line of SN 2017hpa seems to also be stronger than in the comparison SNe~Ia except for SN 2011fe and SN 2012cg.
The pseudo equivalent-width (pEW) of \CII\,$\lambda$6580 is measured to be 14.0$\pm$1.0\,\Angst\ for SN 2017hpa at $t \approx -13$\,d, while those measured for SNe 2005cf, 2011fe and 2012cg are 8.0$\pm$1.1 \Angst, 0.8$\pm$0.2 \Angst\ and 1.0$\pm$0.6 \Angst, respectively. No \CII\ absorption feature is detected in the spectra of SN 2013dy at similar phase. The corresponding pEW of \OI\,$\lambda$7774 is measured as 52.0$\pm$6.3\,\Angst\ for SN 2017hpa at this epoch, comparable to that of SN 2011fe (48.2$\pm$0.4\,\Angst), while those measured for SNe 2012cg and 2013dy are 16.2$\pm$1.4\,\Angst\ and 16.3$\pm$3.4\,\Angst, respectively. No prominent \OI\,$\lambda$7774 is detected in the spectra of SN 2005cf at similar epoch, which is consistent with the findings by \citet{wxf09a}. 
Following the discovery by \citet{zhao16} that the velocity of \OI\,$\lambda$7774 line shows a positive correlation with that of \CII\,$\lambda$6580. We propose that more unburned carbon and oxygen may be kept in the explosion ejecta of SNe 2017hpa, 2011fe, and 2012cg, although the stronger \OI\,$\lambda$7774 absorption observed in SN 2017hpa could result from higher oxygen abundance of the exploding white dwarf \citep{Cui2020}.
A detached \CaII\,NIR HVF could be detected through multi-gaussian fitting proposed by \citet{zhao15,zhao16}.

Figure \ref{fspeccomp}(b) shows the comparison at $\sim 1$\,week before maximum light. All spectra show an increase in absorption strength of IMEs. The \SiII\,$\lambda$6355 velocity of SN 2017hpa derived from absorption minimum is $12,500 \pm 180$\,\kms, which is comparable to that of the comparison sample. The \CII~$\lambda$6580 absorption line remained visible in the red wing of \SiII~$\lambda$6355 at this epoch.

The spectra near maximum light are displayed in Figure \ref{fspeccomp}(c). The absorption features due to IMEs such as \SiII\ at 4130\,\Angst, \SiIII\ at 4560\,\Angst, and \SII\ at 5468, 5612, and 5654\,\Angst\ become prominent at this phase. The \CII\ absorption features at around 6300 and 7000\,\Angst\ remain noticeable in SN 2017hpa but they are barely seen in the comparison SNe~Ia. The $R$(\SiII), defined as the line-strength ratio of \SiII\,$\lambda$5972 to \SiII\,$\lambda$6355 \citep{nug95}, can be used as indicator of the photospheric temperature. A lower value of $R$(\SiII) corresponds to a higher photospheric temperature for the SNe Ia. At around maximum light, $R$(\SiII) is measured to be \RSiII, comparable to that of SN 2018oh ($R$(\SiII) $= 0.15 \pm 0.04$), suggesting that these two SNe have similar photoshperic temperature around maximum light. The relatively larger ratio indicates a lower photospheric temperature for SN 2017hpa. The pseudo-equivalent widths (pEWs) of \SiII\,$\lambda$5972 and \SiII~$\lambda$6355 near maximum light are measured to be $15.5 \pm 0.6$\,\Angst\ and $83.9 \pm 2.2$\,\Angst, respectively, putting SN 2017hpa into the CN subtype of \citet{bran06} classification.

Figure \ref{fspeccomp}(d) shows the spectral evolution at $t \approx 30$\,d. The \CII~$\lambda$6580 absorption line disappeared in this phase in all of our objects. With the receding of the photosphere, the \FeII\ features gain strength and gradually dominate at wavelengths between 4700 and 5000\,\Angst. The absorption profiles of SN 2017hpa and the comparison sample are well developed and tend to have uniform morphologies.

\subsection{Carbon Features}

\begin{figure}[ht]
\centering
\includegraphics[angle=0,width=86mm]{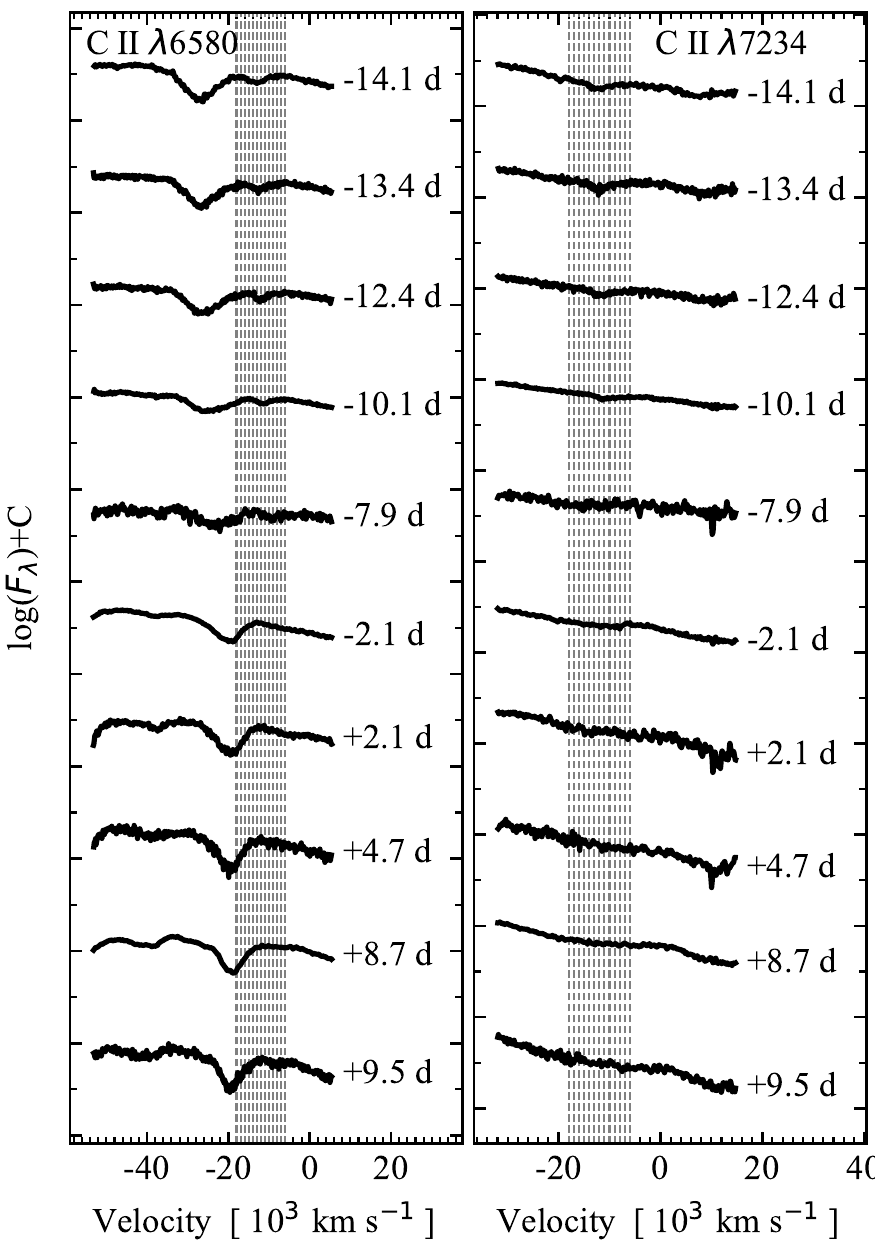}
\caption{The left panel shows the \CII~$\lambda$6580 temporal evolution while the right panel shows that of \CII~$\lambda$7234. The dashed lines mark the Doppler velocity range from $-18,000$\,\kms\ to $-5000$\,\kms. \label{fCIIevolution}}
\end{figure}

The presence of \CII\ absorption can be easily identified in the early-time spectra of SN 2017hpa around 6300 and 7000\,\Angst. The \CII\ absorption features of SN 2017hpa are stronger than in the comparison SNe~Ia. The left panel of Figure \ref{fCIIevolution} shows that the \CII~$\lambda$6580 absorption lines persist in the spectra from $t \approx -14.1$ to $-7.9$\,d. This absorption feature disappeared in the spectra approaching maximum light and then reemerged at $t \approx 9.5$\,d. 
As a possible explanation, we propose that the \CII\ will be highly excited when the detonation front or the deflagration front propagates outward through the ejecta of SNe Ia \citep{Ciaraldi13,Seitenzahl13} and this will make the \CII\ absorption features disappear temporarily. With the receding and cooling of the photosphere of SNe Ia, the \CII\ absorption trough will reemerge in the spectra.
The right panel of Figure \ref{fCIIevolution} shows the relatively weak \CII~$\lambda$7234 absorption; it is noticeable in the earliest four spectra, and it then became flattened. Inspection of the spectra does not reveal significant absorption of \CII~$\lambda$4267 in SN 2017hpa. Both \CII~$\lambda$6580 and \CII~$\lambda$7234 became barely visible in spectra taken $\sim 10$\,d after maximum light.

\begin{figure}[ht]
\centering
\includegraphics[angle=0,width=86mm]{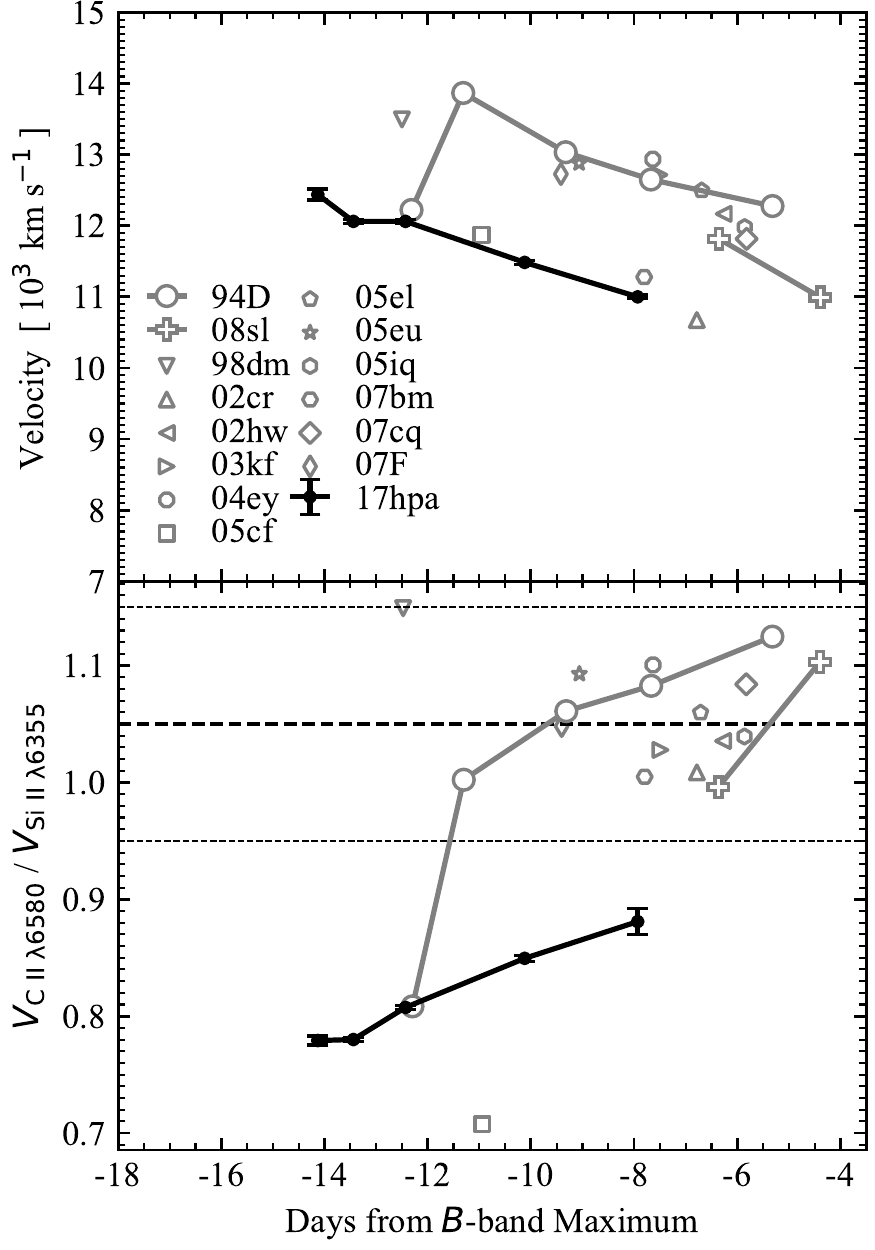}
\caption{Top panel: temporal evolution of the \CII~$\lambda$ expansion velocity for SNe~Ia with carbon detections. Bottom panel: temporal evolution of the velocity ratio \CII~$\lambda$6580 to \SiII~$\lambda$6355. The comparison data are taken from \citet{silver12a}. \label{fCIIRatio}}
\end{figure}

The SN Spectroscopic Evolution\footnote{https://mwvg-spec-evolve.readthedocs.io/en/latest/} package is employed to fit the absorption components of \SiII~$\lambda$6355 and \CII~$\lambda$6580. For SN 2017hpa, the \CII~$\lambda$6580 velocity is found to range from $\sim 13,000$\,\kms\ at $t \approx -14.1$\,d to $\sim 9,300$\,\kms\ at $t \approx -7.9$\,d. According to \citet{silver12a}, the average velocity ratio between \CII~$\lambda$6580 and \SiII~$\lambda$6355 is $\sim 1.05$ for SNe~Ia with observations at least four days prior to $B$-band maximum light. However, the mean  \CII~$\lambda$6580 to \SiII~$\lambda$6355 velocity ratio (hereafter \RCSi) measured for SN 2017hpa is only $\sim 0.81$ (as shown in Figure \ref{fCIIRatio}), suggesting that significant unburned carbon may have mixed deep into the ejecta.

\subsection{Ejecta Velocity}

\begin{figure}[ht]
\centering
\includegraphics[angle=0,width=86mm]{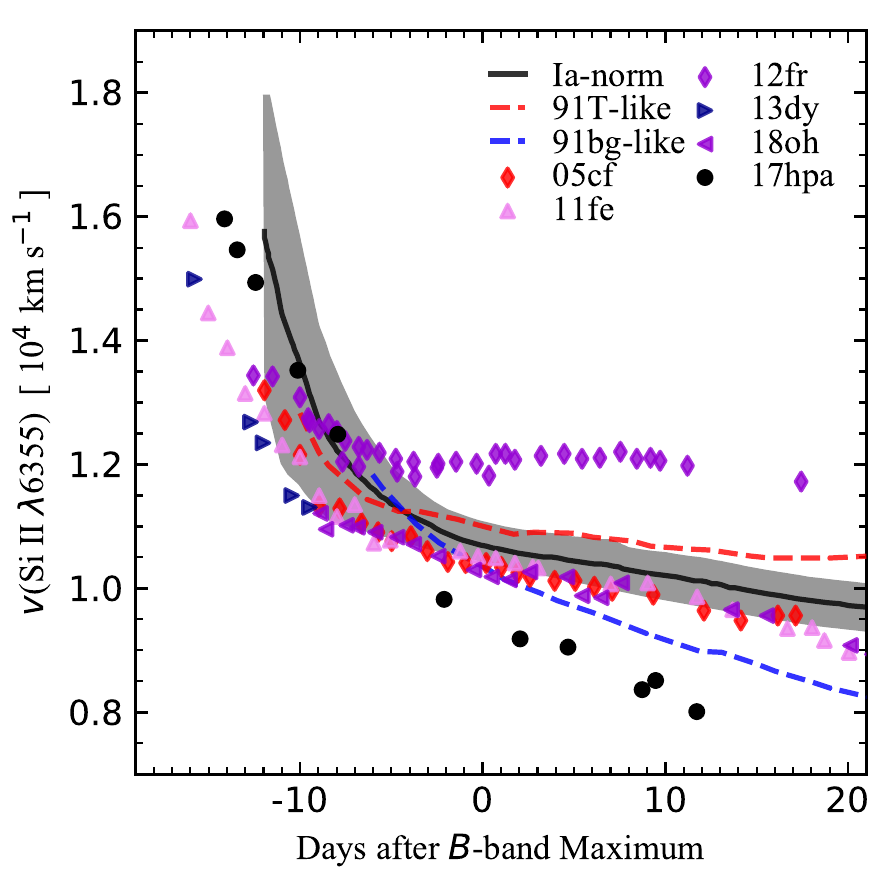}
\caption{Velocity evolution of SN 2017hpa as derived from the absorption minimum of \SiII~$\lambda$6355, compared with SNe 2005cf, 2011fe, 2013dy, and 2018oh. The average velocity curves obtained for SN 1991T-like and SN 1991bg-like SNe are overplotted in red and blue dashed lines, respectively. The normal subclass of SNe~Ia is plotted with a black solid line. The shaded region represents the 1$\sigma$ uncertainty for the mean velocity curve of normal SNe~Ia. Data for the comparison SNe and the region of normal SNe~Ia are extracted from \citep{liwx19}. \label{fSiIIV}}
\end{figure}

\begin{figure}[ht]
\centering
\includegraphics[angle=0,width=86mm]{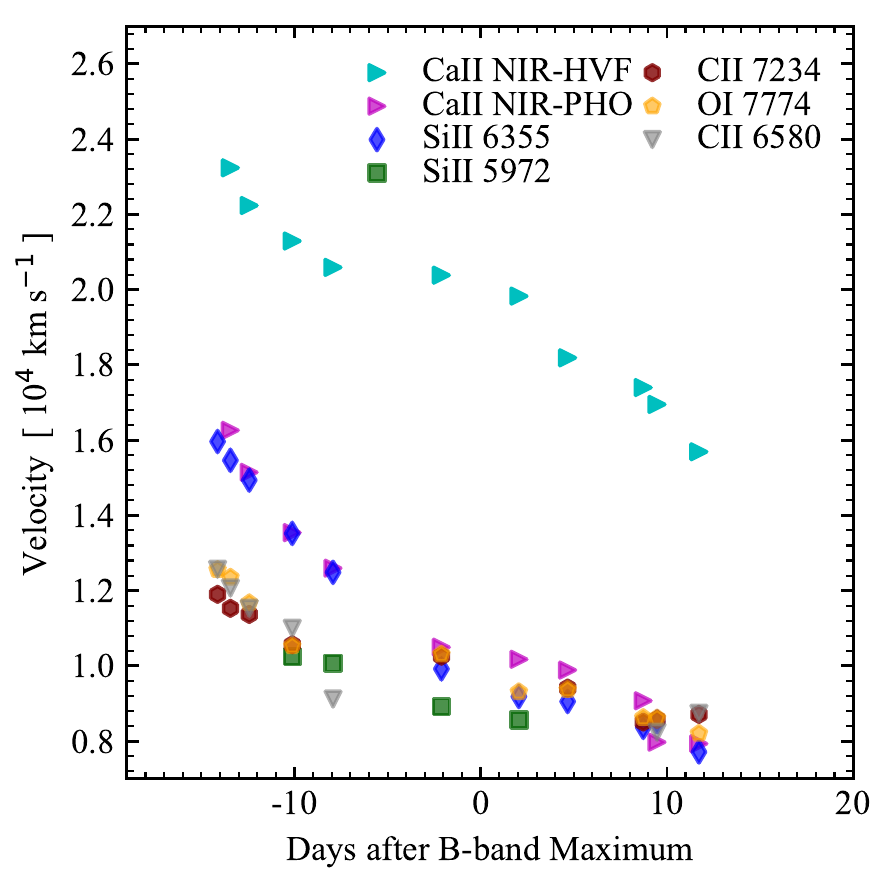}
\caption{Velocity evolution of different elements measured from spectra of SN 2017hpa. \label{felementV}}
\end{figure}

The ejecta velocities measured from the absorption lines, such as \SII~$\lambda\lambda$5468,5640, \SiII~$\lambda$6355, \CII~$\lambda$6580, \CII~$\lambda$7234, and \OI~$\lambda$7774, are shown in Figure \ref{felementV}. The photospheric velocity measured from \SiII~$\lambda$6355 at $t \approx -13.5$\,d is $\sim 16,000$\,\kms, which is comparable to that of the \CaII\ NIR triplet ($\sim 16,200$\,\kms), but it is faster than the \CII\ velocity ($\sim 12,000$\,\kms\ for both \CII~$\lambda$6580 and \CII~$\lambda$7234). The velocity of the \CII~$\lambda$6580 absorption is roughly within the typical expansion velocity of normal SNe~Ia \citep{silver12a}. At the time of $B$-band maximum, the velocity of \SiII~$\lambda$6355 is estimated to be $\sim$~\VSiII, which can put SN 2017hpa into the NV subclass in the \citet{wxf09b} classification scheme (the basic paramaters of SN 2017hpa are listed in Table \ref{tabpars}), as shown in Figure \ref{fSiIIV}. However, the \SiII~$\lambda$6355 velocity of SN 2017hpa seems to have a large gradient, $\sim$ \VSiIIDot, measured within about 10\,d after maximum light.

\begin{table*}
\centering
\caption{Parameters of SN 2017hpa}
\label{tabpars}
\begin{tabular*}{3.0in}{@{\extracolsep{1in}}ll}
\hline\hline
\multicolumn{1}{c}{Parameter} & \multicolumn{1}{c}{Value} \\\hline
\multicolumn{2}{c}{Photometric}                            \\
$B_{\rm max}$                 & \mbobs\,mag  \\
$B_{\rm max}-V_{\rm max}$     & $0.005 \pm 0.007$\,mag \\
$M_{\rm max}(B)$              & \mbmag\,mag \\
$E(B-V)_{\rm host}$           & \ebvalue\,mag   \\
$\Delta m_{15}(B)$            & \dmvalue\,mag \\
$s_{BV}$ 	                   & \sbvalue\     \\
$t_{\rm max}(B)$ 	         & \tbmax\,d \\
$t_0$	                         &	\tzero\,d \\
$\tau_{\rm rise}$	         & \trise\,d \\
$L_{\rm bol}^{\rm max}$	    & \Lmax\ \\
$M_{^{56}\rm Ni}$	          & \MniValue\ \\
\multicolumn{2}{c}{Spectroscopic}                          \\
$v \rm _{0}$(\SiII)	          & \VSiII\ \\
$\dot{v}$(\SiII)		     & \VSiIIDot\ \\
$R$(\SiII)			          & \RSiII\ \\
\hline
\end{tabular*}
\end{table*}

\subsection{High-Velocity Feature}

At early phases, the HVFs of the \CaII\ NIR triplet can be clearly recognized from the coresponding absorption-line profiles in the spectra. We utilize a multi-Gaussian function to fit the absorption profile of the \CaII\ NIR triplet following the method described by \citet{zhao15,zhao16}. For SN 2017hpa, the HVF of the \CaII\ NIR triplet seen in the $t \approx -13.5$\,d spectrum has a velocity of $\sim 24,000$\,\kms, comparable to that of SN 2011fe ($\sim\ 22,000$\,\kms; \citealp{zhangkc16}). The \CaII\ NIR HVFs exhibit a velocity plateau of 20,000\,\kms\ from $t \approx -10$ to $-2$\,d, which is similarly seen in SN 2018oh but at different epochs \citep{liwx19}. Note that there are no obvious \SiII\ HVFs in the early-phase spectra of SN 2017hpa. It is suggested that HVFs are more commonly detected in line profiles of the \CaII\ NIR triplet than in \SiII\ \citep{Maguire2012426,chil14,magu14,Pan446,Silverman451}. Most SNe~Ia are found to have strong \CaII\ NIR HVFs in their spectra at $t < 7$\,d while no more than 30\% of them have strong \SiII\ HVFs \citep{zhao15}.

\section{Discussion}

\subsection{Distance and Quasibolometric Light Curve}

Applying a standard cosmological model and assuming H$_{0} = 73.5$\,\kms\,Mpc$^{-1}$, $\Omega_{M} = 0.3$, and $\Omega_\Lambda = 0.7$ \citep{riess18}, a distance modulus of $\sim 34.05$\,mag can be obtained  for the host galaxy of SN 2017hpa. We also utilize the latest $EBV$ model of SNooPy2 to fit the light curves of SN 2017hpa in several optical bands, and the best-fit result gives an average distance modulus of \distmd\,mag. These two distance moduli agree well with each other within the uncertainties. Adopting the distance modulus as \distmd\,mag and assuming $R_{V} = 3.1$, we derive the absolute $B$-band peak magnitude to be $M_{\rm max}(B)$ = \mbmag\,mag after correcting for both Galactic and host-galaxy extinction. This value agrees well with the typical value of normal SNe~Ia ($M_{\rm max}(B) \approx -19.3$\,mag; \citealp{phi99,wxf09b}).

Our extensive photometric observations are used to establish the quasibolometric light curve of SN 2017hpa. The spectral energy distribution (SED) includes flux contributions from the following bands: $uvw2$, $uvm2$, $uvw1$, $B$, $g$, $V$, $R$, $r$, $I$, and $i$. We adopt the procedure used for SN 2018oh to establish the SED at several epochs \citep{liwx19}. The observed magnitudes are dereddened and converted into flux density. The flux densities are then integrated using Simpson's rule \citep{Rogers20,Syam03} through the effective wavelengths. 


To get better knowledge of the peak luminosity, we use the UV and optical observations to construct the quasibolometric light curves by assuming the NIR contribution to be 5\% at maximum light \citep{lelo09,wxf09a,zhangkc16,zhai16}. Applying a polynomial fitting, the maximum luminosity is estimated to be $L_{\rm peak}$ = \Lmax\ at about 0.85\,d prior to $B$-band maximum. This peak luminosity is comparable to that of SN 2011fe ($\sim 1.13 \times 10^{43}$\,\ergs; \citealp{zhangkc16}) but lower than that of SN 2018oh ($\sim 1.49 \times 10^{43}$\,\ergs; \citealp{liwx19}).

\begin{figure}[ht]
\centering
\includegraphics[angle=0,width=86mm]{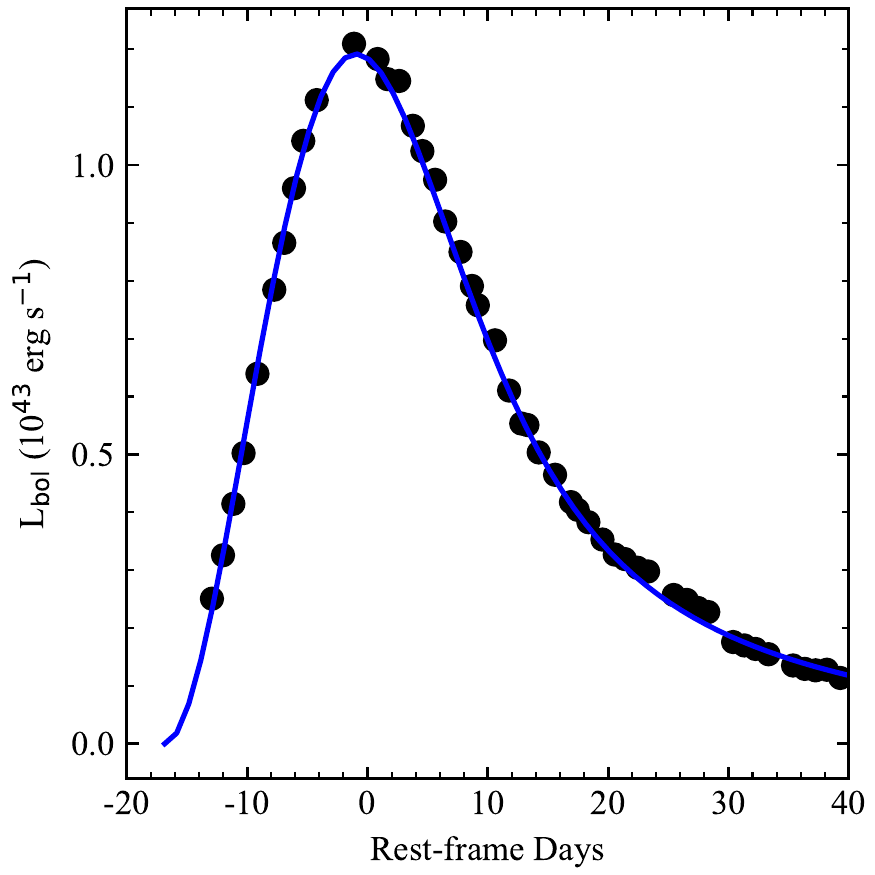}
\caption{The quasibolometric light curve (dots) with an \citet{arn82} radiation diffusion model (blue curve). \label{fbolometric}}
\end{figure}

The modified radiation diffusion model of Arnett \citep{arn82,chat12,liwx19} is applied to evaluate the initial nickel mass together with other physical parameters of the SN ejecta. The Minim Code \citep{chat13} is used to fit the quasibolometric light curve with a constant opacity approximation. The model input parameters are the first-light time (FLT) $t_{0}$, the radioactive \Nifs\ ejecta mass \Mni, the timescale $t_{\rm lc}$ of the light curve, and the leaking timescale of gamma rays $t_{\gamma}$ (see, e.g., \citealp{chat12,chat13}). We set all of these parameters free when performing the model fitting. The final best-fit result of the quasibolometric luminosity evolution of SN 2017hpa is shown in Figure \ref{fbolometric}. Based on $\chi^{2}$ minimization, a set of parameters is found: $t_{0} = -0.94 \pm 1.06$\,d, $t_{\rm lc} =\,$\tlcv\,d, \Mni = \MniValue, and $t_{\gamma} =\,$\tgama\,d. The initial nickel mass is comparable to the estimates of \Mni $\approx 0.57$\,\Msun\ for SN 2011fe \citep{zhangkc16} and $0.55 \pm 0.04$\,\Msun\ for SN 2018oh \citep{liwx19}, but smaller than that of $0.77 \pm 0.11$\,\Msun\ for SN 2005cf \citep{wxf09a} and $0.68 \pm 0.14$\,\Msun\ for SN 2003du \citep{stan07}. 

Adopting the method used by \citet{liwx19}, the average opacity $\kappa$ is estimated to be $0.36 \pm 0.15$\,cm$^2$\,g$^{-1}$. With the best-fit $t_{\rm lc}$ and $t_{\gamma}$, we then obtain the ejecta mass and kinetic energy as \Mej $= 0.70 \pm 0.22$\,\Msun\ and \EK $= (0.70 \pm 0.50) \times 10^{51}$\,erg. These values are within the range of typical SNe~Ia as suggested by \citet{scal19}. 

\subsection{High Velocity Gradient}

\begin{figure*}[ht]
\centering
\includegraphics[angle=0,width=172mm]{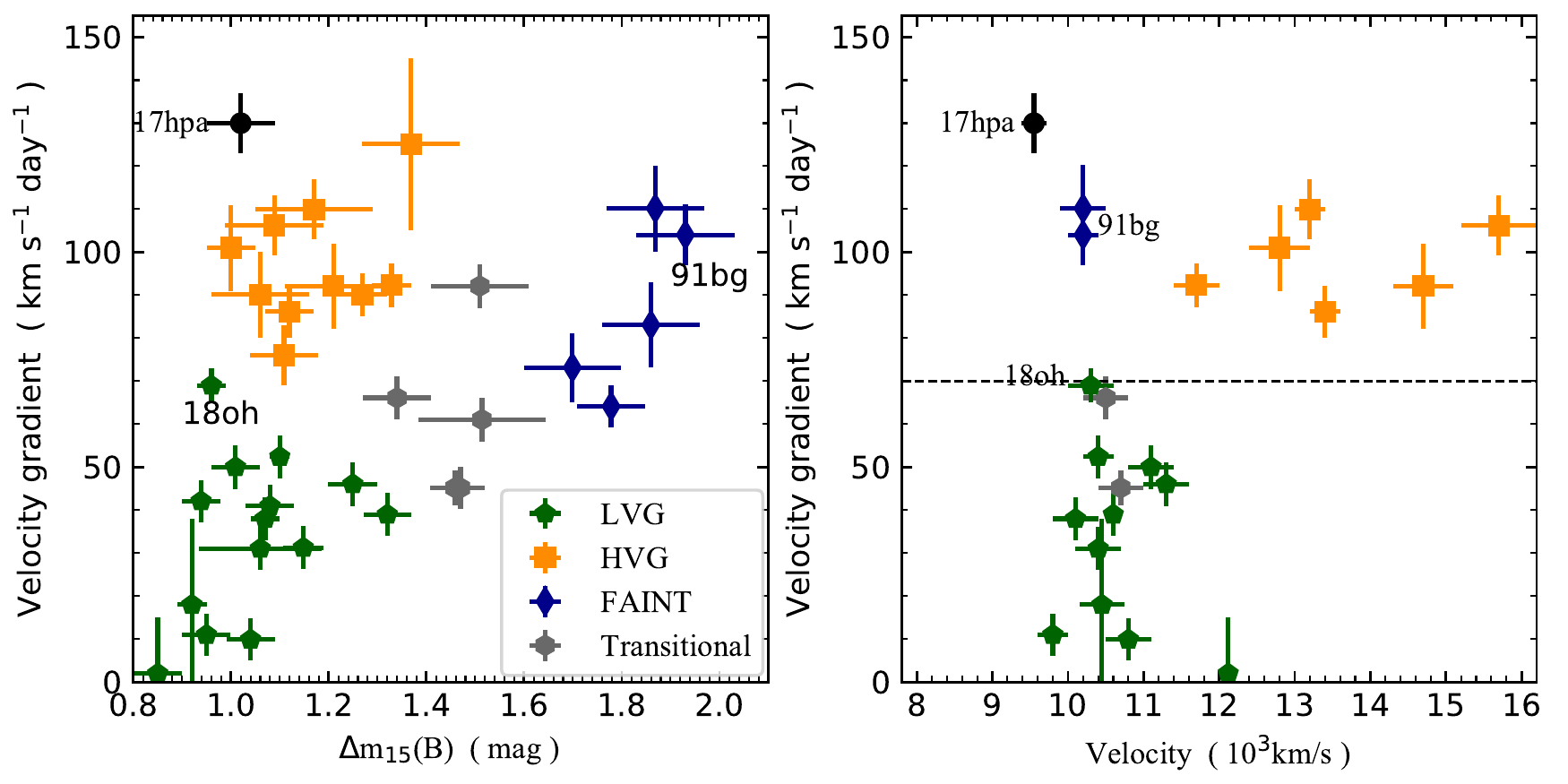}
\caption{Spectroscopic subclassification of SN 2017hpa (as marked with black dot) based on the scheme of \citet{ben05}. Left panel: \mb\ is plotted with respect to the velocity gradient which is measured from \SiII~$\lambda$6355. The SNe from different subtypes are taken from \citet{ben05} and \citet{Chakradhari18}, the four transitional objects are from \citet{past07} and \citet{sahu13}, SN 2005cf is from \citet{wxf09a}, and SN 2018oh is from \citet{liwx19}. Right panel: the scatter plot of the velocity measured from \SiII~$\lambda$6355 near maximum light versus the velocity gradient. The velocities are taken from \citet{silver12c} and \citet{wxf2019}. The horizontal dashed line in the left pannel marks the boundary between HVG and LVG, which is 70\,km\,s$^{-1}$\,d$^{-1}$ \citet{ben05}. \label{HVG}}
\end{figure*}

The ejecta velocity (i.e., \VSiII) measured for SN 2017hpa near maximum light is comparable to that of normal SNe~Ia. According to the velocity gradient of the \SiII, SNe~Ia can be divided into LVG, HVG, and FAINT subtypes \citep{ben05}. Most normal velocity (as opposed to HV) SNe~Ia tend to be LVG or FAINT objects \citep{silver12c}. The left panel of Figure \ref{HVG} shows the scatter plot of the \mb\ versus velocity gradient of SNe~Ia, and the right panel displays that of the velocity gradient vs. velocity  measured around the time of maximum light. It can be seen that SN 2017hpa should be classified in the HVG subcategory, contradicting the trend that SNe~Ia showing prominent carbon features tend to be LVG objects \citep{parr11}. According to previous studies, HV SNe~Ia tend to have larger velocity gradients and vice versa, while this tendency seems to be broken by SN 2017hpa. 

Previous studies have shown that for the HVG and LVG subclasses, the difference in velocity gradient may be due to the different nature of the explosion or the mixing degree of heavy elements \citep{sahu13}. An off-center ignition will trigger SNe to explode asymmetrically. In this case, different viewing angles will cause the observed velocity gradient to vary greatly \citep{Maeda10}. It is suggested that varying the criterion for deflagration-to-detonation transition (DDT; \citealp{Woosley2009}) in explosions also can result in a wide range of velocity gradients \citep{Blondin11}. However, these two scenarios are not suited to SN 2017hpa, which has a low velocity but high velocity gradient. Alternatively, the effective mixing of heavy elements in the SN ejecta may lead to the high velocity gradient of the HVG subclass, while the inefficient mixing may  cause the low velocity gradient of the LVG subclass \citep{Blondin12,sahu13}. 

\subsection{Prominent Carbon Features}

Detection of unburned carbon is important for constraining the explosion mechanisms or progenitor systems of SNe~Ia, and different explosion models predict the presence of unburned material in different regions of the ejecta \citep{fink10,pak12,sim12,seit13,shen18,liwx21}. The velocities of IMEs in DDT will increase with the explosion strength, leading to the diffusion of unburned materials farther outward \citep{fink10,Blondin11}. The carbon and oxygen have distinct velocity distributions in the delayed-detonation model while a similar velocity distribution is predicted in the violent merger \citep{Ropke12}. For those carbon-positive SNe~Ia such as SNe 2005di, 2005el, 2005ki, and SNF20080514-002, the absorption notch due to \CII~$\lambda$6580 usually disappears about one week before maximum light \citep{thomas11}. Recent studies suggest that some SNe Ia or peculiar SNe Ia exhibit unusually persistent carbon features in their spectra, such as SN 2002fk \citep{cart14}, iPTF14atg \citep{cao15}, and SN 2018oh \citep{liwx19}. SN 2017hpa is another example showing such a prominent \CII~$\lambda$6580 feature at early epoches ($t \leq\ -7.9$\,d), and this carbon feature disappeared in near-maximum-light spectra. However, unlike other SNe Ia, the carbon feature seems to reemerge at phases from $t\,\sim\,$+8.7\,d to $\sim\,$+11.7\,d after maximum light.

The velocity ratio of \CII~$\lambda$6580 to \SiII~$\lambda$6355 is an important parameter for setting constraints on the explosion models of SNe~Ia \citep{parr11,fola12}. As shown in Figure \ref{fCIIRatio}, the typical value of such a velocity ratio is $\sim 1.05$ for SNe~Ia \citep{silver12a}. However, the mean \RCSi\ is measured to be 0.81 for SN 2017hpa, much lower than the typical value. 
As noted by \citet{silver12b}, for a given object, the \RCSi\ usually increases somewhat with time. This conclusion may be also supported by the data presented by \citet{parr11}. \citet{Scalzo10} suggested that the prominent \CII~$\lambda$6580 feature, concurrent with low velocities, could be associated with a pre-explosion envelope of progenitor material originated from the merger of two white dwarfs. With the receding of the photosphere of SNe Ia, the ejecta of the inner layer with more uniform velocity distribution begin to show up, which leads to the observed phenomenon that the \RCSi\ increases slowly with time. 
The abundance distribution inferred from the violent merger model indicates that both carbon and oxygen can be mixed  deep into the inner layer of the ejecta \citep{Ropke12}. 
The prominent carbon features and high velocity gradient may suggest that SN 2017hpa is reminiscent of a low-luminosity subclass like SN~1991bg. However, the light curves and color curves of SN 2017hpa are quite different from those of low-luminosity objects like SN~2005bl \citep{taub08} or even transitional objects like SN 2004eo \citep{past07}.

\begin{figure*}[ht]
\centering
\includegraphics[angle=0,width=172mm]{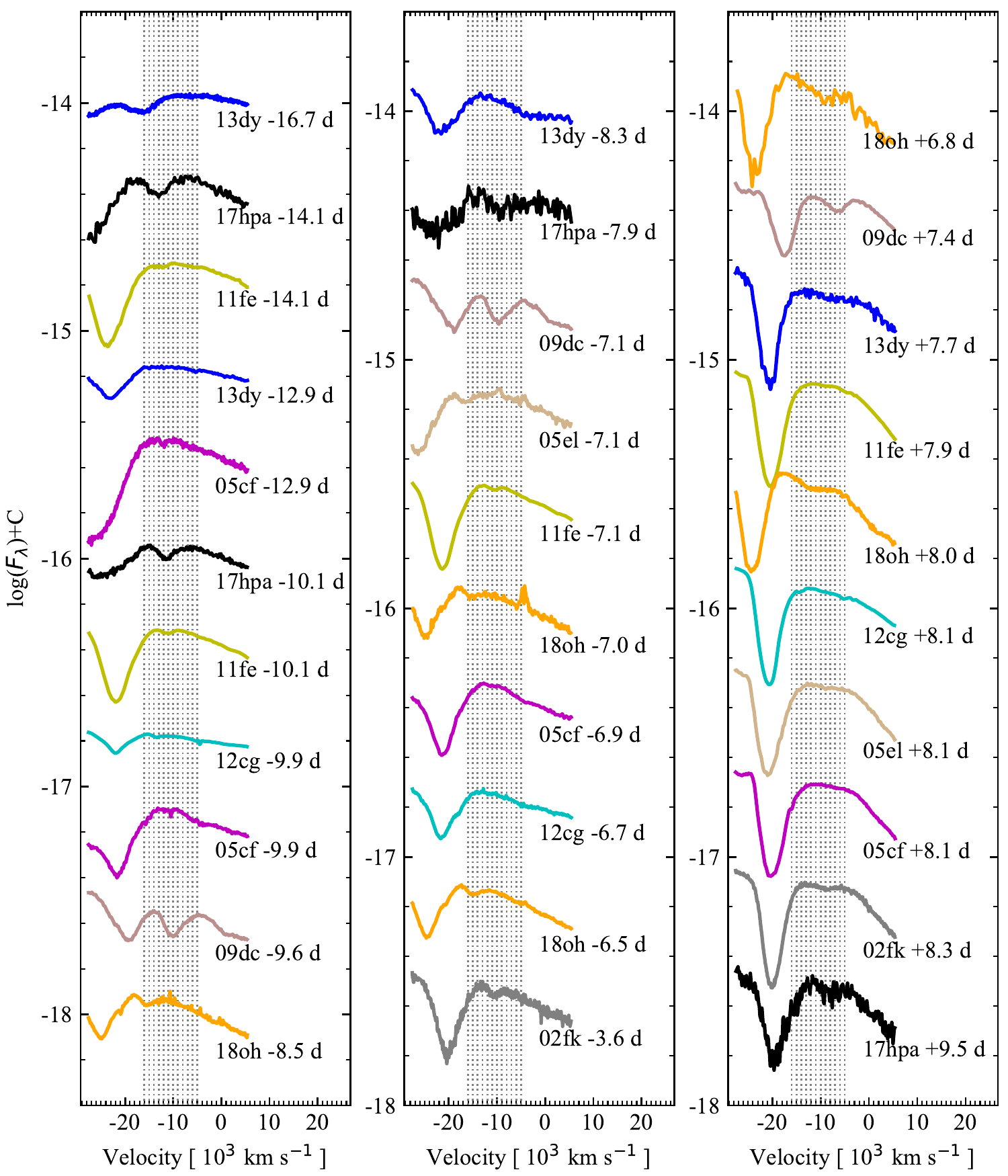}
\caption{The \CII~$\lambda$6580 evolution of SN 2017hpa compared to some well-observed SNe~Ia, including SN 2002fk, 2005el, 2005cf, 2009dc, 2011fe, 2012cg, 2013dy, and 2018oh. \label{fCII6580comp}}
\end{figure*}

To investigate the abnormal behavior of SN 2017hpa, we perform a further comparison of \CII~$\lambda$6580 absorption in Figure \ref{fCII6580comp}, where the sample includes SNe 2002fk, 2005el, 2005cf, 2009dc, 2011fe, 2012cg, 2013dy, and 2018oh. The spectra of the comparison SNe are taken from the \citep{wxf09a,Yaron2012,silver12b,zhai16,zhangkc16,Guillochon2017,liwx19,Stahl2020spec}. In spectra of SN 2013dy, a strong \CII~$\lambda$6580 absorption feature can be found with a velocity up to $\sim 16,300$\,\kms\ at early epochs, but this absorption feature quickly fades away at $t \approx -12.9$\,d, $\sim 3$\,d after explosion \citep{zheng13,pan15}. SN 2012cg shows moderately strong \CII~$\lambda$6580 at the same phase with respect to SN 2017hpa, and the \CII\ absorption feature lasts until $\sim 8$\,d before maximum light \citep{silver12b}. The spectra of SN 2009dc exhibit very prominent \CII~$\lambda$6580 absorption that lasts for a long time, and this SN is proposed to result from a super-Chandrasekhar mass progenitor system \citep{howell2006,sliver11,taub11,tana10}. SNe 2005cf, 2005el, and 2011fe show moderate \CII\ features along their spectra evolution, while SNe 2002fk and 2018oh have \CII\ absorption comparable to that of SN 2017hpa. All three of these normal SNe Ia show prominent \CII\ absorption feature. The \CII\ absorption lines could be detected even in the spectra at $\sim$7\,days after $B$-band maximum light.

Previous studies suggest that carbon-positive SNe~Ia tend to have bluer optical or UV colors \citep{thomas11,silver12b,mil13}. {\it Swift}/UVOT observations also suggest that SNe~Ia with prominent carbon features are NUV-blue objects \citep{rom05, mil10, thomas11}. The only exception is SN 2005cf, which belongs to the NUV-red subgroup but with signatures of carbon features \citep{silver12b,mil13}. SN 2017hpa is also a NUV-red SN~Ia and has carbon in its early-time spectra. Based on model comparisons, \citet{brown19} suggested that the physical origin of the NUV-blue and NUV-red subclasses are likely related to metal abundance. As suggested by \citet{heri17}, the \CII\ absorption features could be hidden by the emission from iron, leading to a lower metallicity within the outer layers of SNe~Ia with carbon signatures. However, if metallicity is the dominant origin of the NUV differences and the presence of \CII\ features, a continuous distribution is expected in each of them \citep{brown19}. A large sample of SNe~Ia with positive \CII\ features is needed for modeling metallicity effects on \CII\ absorption in the spectra.

Based on the above discussions, SN 2017hpa shows prominent carbon features with distinct evolution, a low \CII~$\lambda$6580 to \SiII~$\lambda$6355 velocity ratio, and normal ejecta velocity but a high velocity gradient, all of which are unusual for known subtypes of normal SNe~Ia. We suggest that SN 2017hpa could result from a violent merger explosion of two carbon-oxygen white dwarfs, which brings up the prominent and distinct \CII\ features in its spectra. The deep mixing of the SN ejecta may result in the high velocity gradient of SN 2017hpa. 

\section{Conclusion}

In this paper, we present extensive optical photometry and spectroscopy of the Type Ia SN 2017hpa, which was discovered at a relatively young phase. This object can be put into the category of normal and NUV-red SNe~Ia, with \mb = \dmvalue\,mag and an absolute $B$-band magnitude $M_{\rm max}(B)$ = \mbmag\,mag.

The quasibolometric light curve of SN 2017hpa is established by using extensive UV/optical photometric observations. Arnett's \Nifs\ and \Cofs\ radioactive-decay-driven radiation diffusion model is utilized to fit the quasibolometric light curve, deriving a peak luminosity of SN 2017hpa as $L_{\rm peak}$ = \Lmax. The mass of nickel synthesized during the explosion is estimated to be \Mni = \MniValue, and the ejecta mass \Mej $= 0.70 \pm 0.22$\,\Msun.

The spectral evolution of SN 2017hpa is roughly the same as that of normal SN~Ia such as SN~2018oh. However, prominent \CII\ absorption and abnormal velocity evolution distinguish it from other normal SNe~Ia. The carbon and oxygen features appear stronger than in normal SNe~Ia and lasted until about 10\,d after maximum light, and both the carbon and oxygen have a lower velocity than intermediate-mass elements such as \SiII\ and \CaII. Although SN 2017hpa has a typical ejecta velocity, $\sim$ \VSiII\ as measured near the maximum light, it has an unsually large velocity gradient ($\sim$ \VSiIIDot) in comparison with other normal SNe~Ia. The significant amount of unburned C and O in the ejecta, lower velocity relative to IMEs, and large velocity gradient are more consistent with the merger model. More observations and detailed modeling are needed to reveal the exact explosion physics in objects like SN 2017hpa.

\section*{Acknowledgments}
We thank the anonymous referee for the suggestive comments, which improved the manuscript. Funding for this work was provided by the National Natural Science Foundation of China (NSFC, grants 11873081, U2031209, 12033002, 11633002, and 11761141001) and the National Program on Key Research and Development Project (grant 2016YFA0400803), the High Level Talent-Heaven Lake Program of Xinjiang Uygur Autonomous Region of China. This work is partially supported by the Scholar Program of Beijing Academy of Science and Technology (DZ:BS202002). We acknowledge the staffs of the Lijiang 2.4\,m telescope (LJT), the Xinglong 2.16\,m telescope (XLT), and Lick Observatory for their support. The Chinese Academy of Sciences and the People's Government of Yunnan Province provide support for the LJT, which is corporately run and maintained by Yunnan Observatories and the Center for Astronomical Mega-Science (CAS). JuJia Zhang is supported by the National Natural Science Foundation of China (NSFC; grants 11773067, 11403096), the Youth Innovation Promotion Association of the CAS (grant 2018081), and the  Ten Thousand Talents Program of Yunnan for Top-notch Young Talents. Support for A.V.F.'s group at U.C. Berkeley was provided by the TABASGO Foundation, the Christopher R. Redlich Fund, and the Miller Institute for Basic Research in Science (U.C. Berkeley). 


This work makes use of data from the Las Cumbres Observatory network. JB, DH, DAH, and CP were supported by NSF grant AST-1911225. The {\it Swift}/UVOT data were reduced by P.J. Brown and released in the {\it Swift} Optical/Ultraviolet Supernova Archive (SOUSA), which is supported by NASA's Astrophysics Data Analysis Program (grant NNX13AF35G). Some of the observations with the Lick Observatory 1\,m Nickel telescope were conducted by U.C. Berkeley undergraduate students Sanyum Channa, Edward Falcon,  Nachiket Girish, Romain Hardy, Julia Hestenes, Andrew Hoffman, Evelyn Liu, Shaunak Modak, Costas Soler, Kevin Tang, Sameen Yunus, and Keto Zhang; we thank them for their excellent work. Lick/KAIT and its ongoing operation were made possible by donations from Sun Microsystems, Inc., the Hewlett-Packard Company, AutoScope Corporation, Lick Observatory, the U.S. National Science Foundation, the University of California, the Sylvia \& Jim Katzman Foundation, and the TABASGO Foundation. A major upgrade of the Kast spectrograph on the Shane 3\,m telescope at Lick Observatory was made possible through generous gifts from the Heising-Simons Foundation as well as William and Marina Kast. Research at Lick Observatory is partially supported by a generous gift from Google.

\software{SNooPy \citep{burns11,burns14}, SN-Spectral Evolution (https://github.com/mwvgroup/SN-Spectral-Evolution), Minim Code \citep{chat13}, IRAF \citep{Tody1986,Tody1993}, DAOPHOT \citep{stet87}, PyZOGY \citep{zackay16,guevel17}, lcogtsnpipe \citep{valenti16}, LOSS data-reduction pipeline \citep{gane10,Stahl2019,Stahl2020phot}, Astropy \citep{Astropy2013},
Matplotlib \citep{Hunter07}, Scipy (https://www.scipy.org/), Numpy (https://numpy.org/)}

\clearpage
%

\end{document}